\documentclass[a4paper,12pt]{article}

\usepackage{a4,amsmath,amssymb,dsfont,epsfig}
\usepackage[english]{babel}
\usepackage[hypertex]{hyperref}

\allowdisplaybreaks[4]

\makeatletter

\def\section{\@startsection{section}{1}{\z@}{-3.25ex plus -1ex minus
    -.2ex}{1.5ex plus .2ex}{\normalfont\bfseries}}
\def\subsection{\@startsection{subsection}{1}{\z@}{-3.25ex plus -1ex
    minus -.2ex}{1.5ex plus .2ex}{\normalfont\itshape}}
\@addtoreset{equation}{section}

\renewenvironment{thebibliography}[1]
         {\section*{References}\frenchspacing\small
          \begin{list}{[\arabic{enumi}]}
         {\usecounter{enumi}\parsep=2pt\topsep 0pt
         \settowidth{\labelwidth}{[#1]}
         \leftmargin=\labelwidth\advance\leftmargin\labelsep
         \rightmargin=0pt\itemsep=0pt\sloppy}}{\end{list}}
\makeatother

\newtheorem{theorem}{Theorem}
\newtheorem{lemma}{Lemma}

\newcommand{\les}{\leqslant}
\newcommand{\ges}{\geqslant}
\def\lbt{\left(}
\def\rbt{\right)}

\let\epsilon=\varepsilon

\newcommand{\di}{\genfrac{}{}{0pt}{}}

\begin{document}

\hspace*{\fill}hep-th/0501036

\vskip 15mm

\begin{center}

{\Large\bfseries Renormalization of noncommutative $\phi^4$-theory \\[2mm]
by multi-scale analysis}

\vskip 4ex

Vincent \textsc{Rivasseau}$\,^{1}$, 
Fabien \textsc{Vignes-Tourneret}\,$^{1}$,  
Raimar \textsc{Wulkenhaar}$\,^{2}$

\vskip 3ex  

$^{1}\,$\textit{Laboratore de Physique Th\'eorique, B\^at.\ 210, 
Universit\'e Paris XI \\ F-91405 Orsay Cedex, France}
\\
e-mail: \texttt{vincent.rivasseau@th.u-psud.fr}, 
\texttt{fabien.vignes@th.u-psud.fr}
\\[3ex]
$^{2}\,$\textit{Max-Planck-Institut f\"ur Mathematik in den
  Naturwissenschaften \\Inselstra\ss{}e 22--26, D-04103 Leipzig, Germany}
\\
e-mail: \texttt{raimar.wulkenhaar@mis.mpg.de}
\end{center}

\vskip 5ex

\begin{abstract}
  In this paper we give a much more efficient proof that the real
  Euclidean $\phi^4$-model on the four-dimensional Moyal plane is
  renormalizable to all orders. We prove rigorous bounds on the 
  propagator which complete the previous renormalization proof based
  on renormalization group equations for non-local matrix models.  On
  the other hand, our bounds permit a powerful multi-scale analysis of
  the resulting ribbon graphs. Here, the dual graphs play a particular
  r\^ole because the angular momentum conservation is conveniently
  represented in the dual picture. Choosing a spanning tree in the dual
  graph according to the scale attribution, we prove that the
  summation over the loop angular momenta can be performed at no cost
  so that the power-counting is reduced to the balance of the number
  of propagators versus the number of completely inner vertices in
  subgraphs of the dual graph.
\end{abstract}


\section{Introduction}

Field theories on noncommutative spaces became very popular after the
discovery that they arise in limiting cases of string theory
\cite{Schomerus:1999ug,Seiberg:1999vs}. Although from string theory's
point of view there is no reason that the limit is a well-defined
quantum field theory, there has been an enormous activity aiming at
renormalization proofs for noncommutative quantum field theories. Most
of the attempts focused at the Moyal plane with the associative and
noncommutative product
\begin{align}
  (a\star b)(x) &= \int \frac{d^4k}{(2\pi)^4} \int d^4 y \; a(x{+}\tfrac{1}{2}
  \theta {\cdot} k)\, b(x{+}y)\, \mathrm{e}^{\mathrm{i} k \cdot y}\;.
\label{starprod}
\end{align}
It turned out that the noncommutative analogs of typical field
theoretical (in particular four-dimensional) models on the Moyal plane
are not renormalizable due to the UV/IR-mixing problem
\cite{Minwalla:1999px}.  The construction of dangerous non-planar
graphs was made precise in \cite{Chepelev:2000hm} where the problem
was traced back to divergences in some of the Hepp sectors which
correspond to disconnected ribbon subgraphs wrapping the same
handle of a Riemann surface.

Recently, the renormalization of the noncommutative $\phi^4_4$-model
was achieved \cite{Grosse:2004yu} within a Wilson-Polchinski
renormalization scheme \cite{Wilson:1973jj,Polchinski:1983gv} adapted
to non-local matrix models \cite{Grosse:2003aj}.  The renormalizable
model is defined by the action functional
\begin{align}
  S[\phi] &= \int d^4x \Big( \frac{1}{2} \partial_\mu \phi
  \star \partial^\mu \phi + \frac{\Omega^2}{2} (\tilde{x}_\mu \phi )
  \star (\tilde{x}^\mu \phi ) + \frac{1}{2} \mu_0^2
  \,\phi \star \phi 
+ \frac{\lambda}{4!} \phi \star \phi \star \phi \star
  \phi\Big)(x)\;,
\label{action}
\end{align}
where $\tilde{x}_\mu=2(\theta^{-1})_{\mu\nu} x^\nu$ and the Euclidean
metric is used.

At first sight, the appearance of the translation invariance breaking
harmonic oscillator potential for the $\phi^4$-action (\ref{action})
might appear strange. However, the renormalization proof shows that
there is a marginal interaction which corresponds to that term and as
such requires its inclusion in the initial action. Moreover, thanks to
the oscillator potential, the action (\ref{action}) becomes invariant
under the Langmann-Szabo duality \cite{Langmann:2002cc} which
exchanges position space and momentum space.

We review the main ideas of the renormalization proof, in particular
the analysis of ribbon graphs, in Section~\ref{review}.  However, it
must be underlined that the proof given in \cite{Grosse:2004yu} relies
on a numerical determination of the asymptotic scaling dimensions of
the propagator. Our paper fills this gap by computing rigorous bounds
on the propagator, at least for large enough $\Omega$. This will be
done in Section~\ref{bounds}.

On the other hand, our bounds permit another renormalization strategy which
turns out to be much more efficient. See Section~\ref{sec:powercounting}. The
strategy is inspired by constructive methods \cite{Rivasseau:1991ub}. The key
is a scale decomposition of the propagator and an estimation procedure of the
ribbon graphs which takes into account the scale attribution.  The proof is
carried out for the duals of the ribbon graphs, because the set of independent
variables is particularly transparent in dual graphs.

The methods developed in this paper will be crucial to write a
constructive version of \cite{Grosse:2003aj,Grosse:2004yu}.  Actually
the main obstacle to the construction of the usual $\phi^4$ model is
the non-asymptotic freedom of the theory. In noncommutative
$\mathds{R}^4$, the parameter $\Omega$ controls the UV/IR mixing. When
it reaches $1$, the entanglement is maximum, and the $\beta$ function
vanishes \cite{Grosse:2004by}. In this view, the $\Omega$-region close
to $1$, for which we prove analytical estimates is particularly
important.

\section{Main ideas of the previous renormalization proof}
\label{review}

In order to make this paper self-contained, we review the main ideas
of the renormalization proof given in \cite{Grosse:2004yu} for the
quantum field theory associated with the action (\ref{action}).

In order to avoid the oscillating phase factors of the $\star$-product
in momentum space, the first step is to pass to the matrix base of the
Moyal plane, where the action (\ref{action}) becomes 
\begin{align}
S[\phi]
  &= 4\pi^2 \theta_1 \theta_2 \sum_{m,n,k,l \in \mathds{N}^2} \Big(
  \frac{1}{2} \Delta_{m,n;k,l} \phi_{mn} \phi_{kl} + \frac{\lambda}{4!}  \phi_{mn}
  \phi_{nk} \phi_{kl} \phi_{lm}\Big)\;.
\label{action-matrix}
\end{align}
As usual, we define the quantum field theory by the partition
function, which is expanded into Feynman graphs. 
As the fields are described by matrices $\phi_{mn}$, the resulting
Feynman graphs are ribbon graphs build of propagators and vertices,  
\begin{align}
\parbox{15\unitlength}{\begin{picture}(15,6)
\put(0,2){\epsfig{file=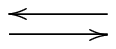,bb=71 667 101 675}}
       \put(1,0){\mbox{\scriptsize$m$}}
       \put(8,-0.5){\mbox{\scriptsize$l$}}
       \put(2,5.5){\mbox{\scriptsize$n$}}
       \put(9,5.5){\mbox{\scriptsize$k$}}
   \end{picture}}
&= G_{m,n;k,l} \;, & 
\parbox{20\unitlength}{\begin{picture}(17,17)
       \put(0,0){\epsfig{file=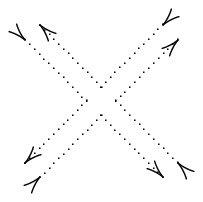,bb=71 638 117 684}}
       \put(-1,5){\mbox{\scriptsize$m_1$}}
       \put(-2,12){\mbox{\scriptsize$n_4$}}
       \put(3,0){\mbox{\scriptsize$n_1$}}
       \put(8,1){\mbox{\scriptsize$m_2$}}
       \put(16,3){\mbox{\scriptsize$n_2$}}
       \put(15,11){\mbox{\scriptsize$m_3$}}
       \put(10,16){\mbox{\scriptsize$n_3$}}
       \put(4,14){\mbox{\scriptsize$m_4$}}
   \end{picture}} &=\delta_{n_1m_2}\delta_{n_2m_3}
\delta_{n_3m_4}\delta_{n_4m_1}\;.
\label{Feynman}
\end{align}
The propagator $G_{mn;kl}$ is the inverse of the kinetic matrix
$\Delta_{mn;kl}$ in (\ref{action-matrix}). We recall the explicit formula in
(\ref{eq:propinit}) and (\ref{eq:propinit-b}). Due to the $SO(2)\times
SO(2)$-symmetry of the action,  $G_{mn;kl} \neq 0$ only if 
$m+k=n+l$. Matrix indices which are not determined by this index
conservation or as external indices of the graph are summation
indices. The corresponding index summation is possibly divergent and
requires a regularization.

In \cite{Grosse:2003aj,Grosse:2004yu} the regularization consists in a
smooth cut-off of the propagator indices as a function of a
renormalization scale $\Lambda$,
\begin{align}
Q_{\di{m^1}{m^2},\di{n^1}{n^2};\di{k^1}{k^2},\di{l^1}{l^2}}(\Lambda) 
= \Lambda \frac{\partial}{\partial \Lambda} 
\left( \prod_{i \in m^1,m^2,\dots,l^1,l^2}
  \chi\Big(\frac{i}{\theta\Lambda^2}\Big) G_{
\di{m^1}{m^2}\di{n^1}{n^2};\di{k^1}{k^2}\di{l^1}{l^2}} \right)\;,  
\label{Q}
\end{align}
where $\chi(x)$ is smooth with $\chi(x)=1$ for $x\les 1$ and $\chi(x)=0$ for
$x \ges 2$. This implies $Q_{mn;kl}(\Lambda) \neq 0$ only if
$\max(m^1,m^2,\dots,l^1,l^2) \in [\theta\Lambda^2,2\theta\Lambda^2]$.  The
graph is then realized by the differentiated cut-off propagators $Q(\Lambda_i)
$ which regulate the index summations. At the end, the nested integral over
$\frac{d\Lambda_i}{\Lambda_i}$ is performed within an interval characterized
by mixed boundary conditions \cite{Polchinski:1983gv}. Actually, the graphs
are build recursively by adding a new propagator. This allows an inductive
proof of the power-counting behavior. On the other hand, one has to carefully
discuss the location of the valence of the graph where one attaches a leg of
the additional propagator. This discussion alone extends over 20 pages in
\cite{Grosse:2003aj}.

It is time for an example. We consider the (planar) one-loop
four-point graph
\begin{align}
&\parbox{38\unitlength}{
\begin{picture}(38,15)
       \put(0,0){\epsfig{file=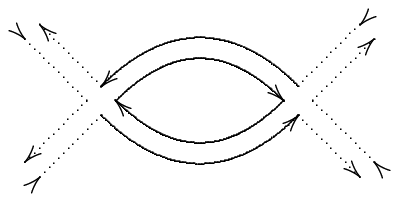,bb=71 630 174 676}}
       \put(-2,12){\mbox{\scriptsize$r$}}
       \put(1,5){\mbox{\scriptsize$r$}}
       \put(34,10){\mbox{\scriptsize$s$}}
       \put(36,3){\mbox{\scriptsize$s$}}
       \put(4,0){\mbox{\scriptsize$n$}}
       \put(30,0){\mbox{\scriptsize$k$}}
       \put(5,14){\mbox{\scriptsize$m$}}
       \put(31,15){\mbox{\scriptsize$l$}}
       \put(11.5,8){\mbox{\scriptsize$p{+}m$}}
       \put(19,7){\mbox{\scriptsize$p{+}l$}}
   \end{picture}}
\nonumber
\\*[2ex]
&= \left\{\int_{\Lambda_R}^\Lambda \frac{d\Lambda_2}{\Lambda_2} 
\int_{\Lambda_2}^{\Lambda_0} \frac{d\Lambda_1}{\Lambda_1} 
\sum_{p} Q_{m,p+m;p+l,l}(\Lambda_2)\,
Q_{n,p+m;p+l,k}(\Lambda_1) \right\} + \{
\Lambda_1 \leftrightarrow \Lambda_2\}
\nonumber
\\*
& +A_{rm;ls;sk;nr}(\Lambda_R)\;.
\label{4point}
\end{align}
The choice of the boundary conditions is a preliminary one which
ensures \emph{convergent integrals at the expense of infinitely many initial
data} $A_{rm;ls;sk;nr}(\Lambda_R)$. This will be corrected later in
(\ref{4point-i}). 

One of the bounds we prove in this paper can be put in the following
form
\begin{align}
G_{\di{m^1}{m^2},\di{n^1}{n^2};\di{k^1}{k^2},\di{l^1}{l^2}} 
\les  K \int_0^1 d\alpha \,e^{-
c \alpha (m^1+m^2+n^1+n^2+ k^1+k^2+ l^1+l^2)}\;.
\label{est-Delta}
\end{align}
\phantom{a}From the remarks made after (\ref{Q}) on the range of the maximal index we
conclude
\begin{align}
\left|Q_{\di{m^1}{m^2},\di{n^1}{n^2};\di{k^1}{k^2},\di{l^1}{l^2}}
  (\Lambda) \right|
&\les 32 K \max_x \chi'(x) \int_0^1 d\alpha\; 
e^{- c \theta\Lambda^2 \alpha} 
\les \frac{32 K \max_x \chi'(x)}{c \theta\Lambda^2 }\;.
\label{est-Q}
\end{align}
We thus estimate the summation over $p$ in (\ref{4point}) by
the maximum of the propagators $Q$ over $p$ and a volume factor
$(2\theta \Lambda_2^2)^2$ from the support of the cut-off
function. This shows that the integral (\ref{4point}) is estimated by
a constant times $\ln \frac{\Lambda}{\Lambda_R}$. 

The scaling of (\ref{est-Q}) and the volume of the support of
(\ref{Q}) with respect to any index seem to suggest that $N$-point
graphs have, as in commutative $\phi^4_4$-theory, a power-counting
degree $4-N$. However, this conclusion is too early.  Namely, there is
a problem in presence of completely inner vertices, which require
additional index summations. The following graph
\begin{align}
\parbox{50\unitlength}{\begin{picture}(50,32)
       \put(0,0){\epsfig{file=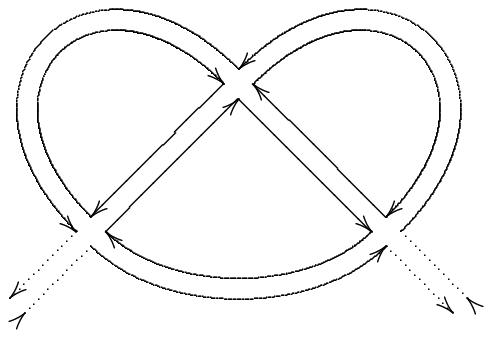,bb=92 571 226 661}}
       \put(0,5){\mbox{\scriptsize$m$}}
       \put(3,0){\mbox{\scriptsize$n$}}
       \put(44,6){\mbox{\scriptsize$l$}}
       \put(40,2){\mbox{\scriptsize$k$}}
       \put(24,29){\mbox{\scriptsize$q$}}
       \put(6,15){\mbox{\scriptsize$p_1{+}m$}}
       \put(14,23){\mbox{\scriptsize$p_1{+}q$}}
       \put(35,15){\mbox{\scriptsize$p_2{+}l$}}
       \put(27,23){\mbox{\scriptsize$p_2{+}q$}}
       \put(20,17){\mbox{\scriptsize$p_3{+}q$}}
       \put(29,8){\mbox{\scriptsize$p_3{+}l$}}
       \put(12,9){\mbox{\scriptsize$p_3{+}m$}}
\end{picture}}
\label{graph-at2}
\end{align}
entails \emph{four} independent summation indices $p_1,p_2,p_3$ and
$q$, whereas for the power-counting degree $4-N$ we should only have
three of them. It requires a more careful analysis of the scaling
behavior of the propagator to show that the $q$-summation can
actually be performed at no cost, i.e.\ without a volume factor. The
reason is that the propagators show some sort of quasi-locality which
implies that the contribution of a propagator $G_{m,n;k,l}$ to a
graph is strongly suppressed if $\|m-l\|$ is large. Thus, 
taking for given $m$ the entire sum over $l$ does not change the
power-counting behavior, 
\begin{align}
\left|\max_{m^i} \left(\sum_{l^1,l^2} \max_{n^i,k^i} 
Q_{\di{m^1}{m^2},\di{n^1}{n^2};\di{k^1}{k^2},\di{l^1}{l^2}}
  (\Lambda) \right)\right|
&\les \frac{K'}{\theta\Lambda^2 }\;.
\label{est-Q1}
\end{align}

The two bounds (\ref{est-Q}) and (\ref{est-Q1}) together ensure the
expected power-counting behavior for all \emph{planar} ribbon
graphs. But (\ref{est-Q1}) does even more: it ensures the
\emph{irrelevance of all non-planar graphs}. For instance, in the
non-planar graphs
\begin{align}
&\left.\parbox{43mm}{\begin{picture}(20,20)
       \put(0,0){\epsfig{file=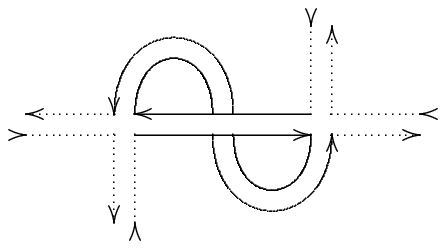,bb=71 625 187 684}}
       \put(2,13){\mbox{\scriptsize$m_4$}}
       \put(0,7){\mbox{\scriptsize$n_4$}}
       \put(4,2){\mbox{\scriptsize$m_1$}}
       \put(13,0){\mbox{\scriptsize$n_1$}}
       \put(36,13){\mbox{\scriptsize$n_2$}}
       \put(34,7){\mbox{\scriptsize$m_2$}}
       \put(32,18){\mbox{\scriptsize$m_3$}}
       \put(25,20){\mbox{\scriptsize$n_3$}}
       \put(13.5,13.5){\mbox{\scriptsize$q$}}
       \put(25,6.5){\mbox{\scriptsize$q'$}}
   \end{picture}} \right|_{q'=n_1+n_3-q}~
\left.\parbox{41mm}{\begin{picture}(20,24)
       \put(0,0){\epsfig{file=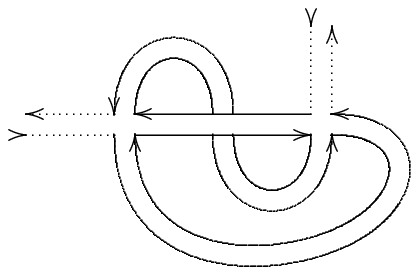,bb=71 613 184 684}}
       \put(2,17){\mbox{\scriptsize$m_2$}}
       \put(0,11){\mbox{\scriptsize$n_2$}}
       \put(13,11){\mbox{\scriptsize$r'$}}
       \put(32,11){\mbox{\scriptsize$r$}}
       \put(32,22){\mbox{\scriptsize$m_1$}}
       \put(25,24){\mbox{\scriptsize$n_1$}}
       \put(13.5,17.5){\mbox{\scriptsize$q$}}
       \put(25,10.5){\mbox{\scriptsize$q'$}}
   \end{picture}} \right|_{\mbox{\scriptsize$\begin{array}{l}
q'=m_2+r-q \\ r'=n_2+r-m_1 \end{array}$}} \hspace*{-1em}
\label{np-graphs}
\end{align}
the summation over $q$ and $q,r$, respectively, is controlled by
(\ref{est-Q1}), i.e.\ the quasi-locality of the propagator, so that
the graphs in (\ref{np-graphs}) can be estimated without any volume
factor. 

We recall from \cite{Grosse:2003aj} that the non-planarity of ribbon
graphs is classified by the number $B$ of boundary components and the
genus $g=1-\frac{1}{2}(F-I+V)$ of the Riemann surface on which the
graph is drawn. Here, $V$ and $I$ are the number of vertices and edges
(inner double lines) of the graph. To determine the number $F$ of
faces we close the external legs, that is, we connect the outgoing
arrow labelled $m_i$ of an external leg directly with its incoming
arrow $n_i$. Then, $F$ is the number of closed single lines and $B$
the number of those closed lines which carry external legs. Then,
according to \cite{Grosse:2003aj, Grosse:2004yu}, the power-counting
degree of a $N$-leg ribbon graph in four dimensions is 
\begin{align}
\omega= (4-N) - 4(2g+B-1)\;.
\end{align}
The left graph in (\ref{np-graphs}) has topology $B=2, g=0$ and the
right graph $B=1,g=1$.

As a result, there remain only the planar two- and four-leg graphs
which can be relevant and marginal. The quasi-locality of the
propagator improves the situation in selecting only 
\begin{itemize}
\item the planar four-leg graphs with \emph{constant index} along the
  trajectory as marginal,

\item the planar two-leg graphs with \emph{constant index} along the
  trajectory as relevant,

\item the planar two-leg graphs with an \emph{accumulated index jump
    of 2} along the trajectory as marginal.
\end{itemize}
We refer to \cite{Grosse:2004yu} for details. The trajectories are
the open single lines of the graph (before the closure which
identifies the faces). This leaves still an infinite number of
divergent graphs. However, there is a discrete Taylor expansion about
vanishing external indices which decomposes these divergent graphs 
into four relevant and marginal base functions and an irrelevant
remainder. For instance, the decomposition for the marginal case $m=l$
and $n=k$ of the graph (\ref{4point}) reads 
\begin{align}
&\parbox{38\unitlength}{
\begin{picture}(38,15)
       \put(0,0){\epsfig{file=a24,bb=71 630 174 676}}
       \put(-2,12){\mbox{\scriptsize$r$}}
       \put(1,5){\mbox{\scriptsize$r$}}
       \put(34,10){\mbox{\scriptsize$s$}}
       \put(36,3){\mbox{\scriptsize$s$}}
       \put(4,0){\mbox{\scriptsize$n$}}
       \put(30,0){\mbox{\scriptsize$n$}}
       \put(5,14){\mbox{\scriptsize$m$}}
       \put(30,15){\mbox{\scriptsize$m$}}
       \put(11.5,8){\mbox{\scriptsize$p$}}
       \put(23,7){\mbox{\scriptsize$p$}}
   \end{picture}}
\nonumber
\\*[2ex]
&= \Bigg\{\int_{\Lambda}^{\Lambda_0} \frac{d\Lambda_2}{\Lambda_2} 
\int_{\Lambda_2}^{\Lambda_0} \frac{d\Lambda_1}{\Lambda_1} 
\sum_{p} \big(
Q_{m,p;p,l}(\Lambda_2)\,Q_{n,p;p,n}(\Lambda_1) 
-Q_{0,p;p,0}(\Lambda_2)\,Q_{0,p;p,0}(\Lambda_1) \big) 
\nonumber
\\
&+ \int_{\Lambda_R}^{\Lambda} \frac{d\Lambda_2}{\Lambda_2} 
\int_{\Lambda_2}^{\Lambda_0} \frac{d\Lambda_1}{\Lambda_1} 
\sum_{p} 
Q_{0,p;p,0}(\Lambda_2)\,Q_{0,p;p,0}(\Lambda_1) \Bigg\}
+ \{
\Lambda_1 \leftrightarrow \Lambda_2\} +A_{00;00;00;00;00}(\Lambda_R)\;.
\label{4point-i}
\end{align}
Thus, this definition necessitates a single initial value
$A_{00;00;00;00;00}(\Lambda_R)$ which represents the
normalization condition for the coupling constant. 

Of particular importance are the marginal two-leg graphs with an
accumulated index jump by 2, such as 
\begin{align}
\sum_{p,p',q,q'}~
\parbox{52mm}{\begin{picture}(20,27)
       \put(0,0){\epsfig{file=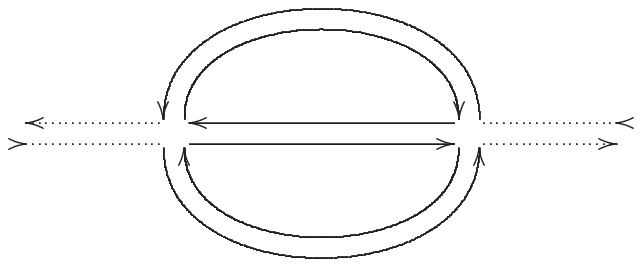,bb=71 603 243 678}}
       \put(10,17){\mbox{\scriptsize$\di{1}{0}$}}
       \put(48,17){\mbox{\scriptsize$\di{0}{0}$}}
       \put(10,9){\mbox{\scriptsize$\di{1}{0}$}}
       \put(48,9){\mbox{\scriptsize$\di{0}{0}$}}
       \put(20,9){\mbox{\scriptsize$\di{q+1}{q'}$}}
       \put(20,17){\mbox{\scriptsize$\di{p+1}{p'}$}}
       \put(38,9){\mbox{\scriptsize$\di{q}{q'}$}}
       \put(38,17){\mbox{\scriptsize$\di{p}{p'}$}}
     \end{picture}}
\end{align}
The corresponding initial value represents the normalization condition
for the frequency parameter $\Omega$ in the initial action
(\ref{action}). Therefore, the harmonic oscillator potential must be
present from the beginning in order to obtain a renormalizable model.


\section{Bounds for the propagator}
\label{bounds}

\subsection{Propagator in the matrix base and cut-offs}

The propagator of the noncommutative $\phi^4$-model in the matrix base
of the $D$-dimensional Moyal plane is given by\footnote{Our
  representation (\ref{eq:propinit}) and (\ref{eq:propinit-b})
  corresponds to (A.17) in \cite{Grosse:2004yu} with $z=1-\alpha$. The
  often used index parameter $\alpha$ in \cite{Grosse:2004yu} is
  denoted by $h$.} a positive sum \cite{Grosse:2004yu}, analogous to
the heat-kernel or parametric $\alpha$-space representation
$\frac{1}{p^{2}+m^{2}}=\int_{0}^{\infty}d\alpha\,
e^{-\alpha(p^{2}+m^{2})}$ of the ordinary commutative propagator:
\begin{align}
  \label{eq:propinit}
  G_{m, m+h; l + h, l} 
&= \frac{\theta}{8\Omega} \int_0^1 d\alpha\,  
\dfrac{(1-\alpha)^{\frac{\mu_0^2 \theta}{8 \Omega}+(\frac{D}{4}-1)}}{  
(1 + C\alpha )^{\frac{D}{2}}} \prod_{s=1}^{\frac{D}{2}} 
G^{(\alpha)}_{m^s, m^s+h^s; l^s + h^s, l^s}\;,
\\
 G^{(\alpha)}_{m, m+h; l + h, l}
&= \lbt\frac{\sqrt{1-\alpha}}{1+C \alpha} 
\rbt^{m+l+h} \sum_{u=\max(0,-h)}^{\min(m,l)}
   {\cal A}(m,l,h,u)\ 
\lbt \frac{C \alpha (1+\Omega)}{\sqrt{1-\alpha}\,(1-\Omega)} 
\rbt^{m+l-2u}\;,
\label{eq:propinit-b}
\end{align}
where ${\cal A}(m,l,h,u)=\sqrt{\binom{m}{m-u}
\binom{m+h}{m-u}\binom{l}{l-u}\binom{l+h}{l-u}}$ and $C$ is a function
of $\Omega$, namely $C(\Omega)=
\frac{(1-\Omega)^2}{4\Omega}$.  Indices such as $m,l,h$ and $u$ have
$\frac{D}{2}$ non-negative components $m^s,l^s,h^s,u^s$, one for each
symplectic pair of $\mathds{R}^D$. However, due to (\ref{eq:propinit})
it is enough to prove estimations for a single component. We define
the norm of an index by $\|m\|=\sum_{s=1}^{D/2} m^s$.

We know that cut-offs in the parametric representation for
\emph{commutative} theories are specially convenient both for
perturbative and constructive renormalization. In the same spirit we
will divide the integral (\ref{eq:propinit}) into slices. First we
divide it into two different regions
\begin{itemize}
\item $M^{-1}\les \alpha \les 1$ where we expect an exponential decay
  in $m+l+h$ of order ${\cal O}$(1),
\item $0 \les \alpha \les M^{-1}$. This is the UV/IR region which is
  further sliced according to a geometric progression. For each slice
  we expect a scaled exponential decay.
\end{itemize}
The real number $M>1$ has a carefully chosen $\Omega$-dependence. Then, the
decomposition
\begin{equation}
  \label{eq:slices}
  \int_{0}^1 d\alpha = \sum_{i=1}^\infty \int_{M^{-i}}^{M^{-i+1}} 
d\alpha
\end{equation}
leads to the following propagator for the $i^{\text{th}}$ slice: 
\begin{align}
G^i_{m,m+h,l+h,l} 
&=\frac{\theta}{8\Omega}  \int_{M^{-i}}^{M^{-i+1}} d\alpha\; 
\dfrac{(1-\alpha)^{\frac{\mu_0^2 \theta}{8 \Omega}+(\frac{D}{4}-1)}}{  
(1 + C\alpha )^{\frac{D}{2}}} \prod_{s=1}^{\frac{D}{2}} 
G^{(\alpha)}_{m^s, m^s+h^s; l^s + h^s, l^s}\;.
\label{prop-slice-i}
\end{align}
The first slice $i=1$ is treated separately.

Remark that the factor ${\cal A}$ in (\ref{eq:propinit-b}) is the only
one which prevents us from explicitly performing the $u$-sum. All the bounds
in this paper are obtained by applying to the binomial coefficients in
${\cal A}$ the simple overestimate $\binom{n}{q}\les\frac{n^{q}}{q!}$.
Of course, this bound is sharp only for $q\ll n$. In the regime $n-q\ll
n$ one should rather use the symmetric bound
$\binom{n}{q}\les\frac{n^{n-q}}{(n-q)!}$. 

For $\alpha=0$ we see from (\ref{eq:propinit-b}) that the propagator vanishes
unless $u=l=m$. This suggests to bound ${\cal A}$ by
\begin{equation}
  \label{eq:Auvir}
\mathcal{A}(m,l,h,u) 
\les \frac{\sqrt{m(h +m)}^{m-u}\sqrt{l(h +l)}^{l-u} }{(m-u)!(l-u)!}
\les \frac{(m+h/2)^{m-u} (l+h/2)^{l-u} }{(m-u)!(l-u)!}\;.
\end{equation}
Hence, for $\alpha \les M^{-1}$,
\begin{align}
G^{(\alpha)}_{m,m+h;l+h,l} 
&\les \left(\frac{4\Omega\sqrt{1-\alpha}}{4\Omega + (1-\Omega)^2\alpha}
\right)^{m+l +h} 
\nonumber
\\*
&\times 
\sum_{u=\max(0,-h)}^{\min(m,l)} \!\!\!
  \frac{\displaystyle  \left( \frac{\alpha(1-\Omega^2)
\sqrt{m(m+h)}}{4\Omega \sqrt{1-\alpha}} \right)^{\!m-u}}{(m-u)!} 
\frac{\displaystyle 
\left( \frac{\alpha(1-\Omega^2)\sqrt{l(l+h)}}{
4\Omega \sqrt{1-\alpha}}  \right)^{\!l-u} }{(l-u)!}\,.
\label{inijump}
\end{align}

On the other hand, we observe from (\ref{eq:propinit-b}) that for $\alpha =1$
the propagator vanishes unless $u=h=0$. In this situation the bound
(\ref{eq:Auvir}) is not suitable anymore because $m-u$ and $l-u$ are of order
$\mathcal{O}(m)$ and $\mathcal{O}(l)$, respectively. Instead, we can use
\begin{equation}
  \label{eq:A0a}
\mathcal{A}(m,l,h,u)
\les\frac{\sqrt{ml}^u\sqrt{(m+h)(l+h)}^{h+u}}{u!(h+u)!}%
  \les\frac{((m+l)/2)^u((m+l+2h)/2)^{h+u}}{u!(h+u)!}\;.
\end{equation}
Inserting (\ref{eq:A0a}) into (\ref{eq:propinit-b}) we obtain
\begin{align}
G^{(\alpha)}_{m, m+h; l + h, l}
&\les \left(\frac{\alpha(1-\Omega^2)}{4\Omega+(1-\Omega)^2\alpha} 
\right)^{m+l+h} 
\nonumber
\\*
& \times \sum_{u=\max(0,-h)}^{\min(m,l)} \hspace*{-0.5em}
\dfrac{ 
\left(\dfrac{4\Omega\sqrt{1-\alpha}\sqrt{ml}}{\alpha (1-\Omega^2)}
\right)^u}{u!}
\dfrac{ 
\left(\dfrac{4\Omega\sqrt{1-\alpha}\sqrt{(m+h)(l+h)}}{\alpha (1-\Omega^2)}
\right)^{u+h}}{(u+h)!}\;.
\label{inijump-2}
\end{align}

The further procedure will be to use the estimation
\begin{equation} \label{eq:exp1}
  \sum_{u=0}^{\min (m,l)}
\ \frac{X^{m-u}}{(m-u)!} 
\frac{Y^{l-u}}{(l-u)!}
\les e^{X+Y}\;.
\end{equation}
Then, we have to find conditions on $\Omega$ and $\alpha$ under which a
certain exponent is negative. These conditions are given in the following
Lemma:
\begin{lemma}
\label{Lemma1}
Let 
\begin{align}
R(\Omega):=1-\frac{9}{10}\left(
\left(\frac{1-\Omega}{1+\Omega}\right)  
\ln \left(\frac{1-\Omega}{1+\Omega}\right)\right)^2
\label{Rbeta}
\end{align}
and $\Omega_R$ be the position of the maximum of $R(\Omega)$, i.e.\ 
$R'(\Omega_R) =0$. One has approximately $\Omega_R=0.462117$. 
We define 
\begin{align}
M^{-1}= \left\{ \begin{array}{ll} 
R(\Omega) & \text{for } \Omega \ges \Omega_R \;,\\  
R(\Omega_R) \qquad &  
\text{for } \Omega \les \Omega_R\;.
\end{array}\right.
\label{M-1}
\end{align}
Then, for all $\alpha \in [M^{-1},1]$ and all 
$\Omega \in (0,1)$ one has 
\begin{align}
E(\Omega,\alpha) 
:= \dfrac{4\Omega\sqrt{1-\alpha}}{\alpha (1-\Omega^2)}
+ \ln \left(\frac{\alpha(1-\Omega^2)}{4\Omega+(1-\Omega)^2\alpha}\right)
\les - \frac{1}{15}\Omega M^{-1}\;.
\label{eq-lem1}
\end{align}
\end{lemma}
{\bf Proof.} We have
\begin{align}
\frac{\partial}{\partial \alpha} E(\Omega,\alpha) 
&= \frac{2\Omega}{(1-\Omega^2)\alpha^2\sqrt{1-\alpha}}
\Big((\alpha-2)+ \frac{2(1-\Omega^2)\alpha\sqrt{1-\alpha}}{
4\Omega+(1-\Omega)^2 \alpha}\Big)
\nonumber
\\*
&\les \frac{2\Omega}{(1-\Omega^2)\alpha^2\sqrt{1-\alpha}}
\Big((\alpha-2)+ \frac{(1-\Omega^2)\alpha(2-\alpha)}{
4\Omega+(1-\Omega)^2 \alpha}\Big)
\nonumber
\\*
&=-\frac{2\Omega(2-\alpha)(2\Omega^2\alpha+(4-2\alpha)\Omega)}{
(1-\Omega^2)\alpha^2\sqrt{1-\alpha}
(4\Omega+(1-\Omega)^2 \alpha)}<0\;.
\label{E-deriv}
\end{align}
Thus, the function $E(\Omega,\alpha)$ is
monotonously decreasing in $\alpha$ and, comparing $E(\Omega,1)=
\ln\frac{1-\Omega}{1+\Omega}<0$ with $E(\Omega,0)=+\infty$, has a single
zero $E(\Omega,\alpha_0)=0$ with $\alpha_0 \approx 1$. Developing
$E(\Omega,\alpha)$ about $\alpha=1$, the leading term is of order
$\frac{1}{2}$:
\begin{align}
E(\Omega,\alpha) 
:= \sqrt{1-\alpha}\,\frac{4\Omega}{(1+\Omega)^2} 
\frac{(1+\Omega)}{(1-\Omega)} + \ln\frac{1-\Omega}{1+\Omega}
+ \mathcal{O}(1-\alpha)\;.
\end{align}
This means that for $\Omega\approx 1$, the zero is found near
$\alpha_0=1-\left(\frac{(1+\Omega)^2} {4\Omega}
\frac{(1-\Omega)}{(1+\Omega)}\ln\frac{1-\Omega}{1+\Omega}\right)^2$. We
know
from (\ref{E-deriv}) that $E(\Omega,\alpha)<0$ for $\alpha \in (\alpha_0,1]$.
To be on the safe side with respect to higher order terms, we make the above
estimation for $\alpha_0$ slightly bigger by removing the factor 
$\frac{(1+\Omega)^2}{4\Omega} \ges 1$ and by rescaling the result 
by $\frac{9}{10}$. This leads to (\ref{Rbeta}). We can now plot the function 
$E(\Omega,M^{-1})$ over $\Omega$ and compare it with $-\frac{1}{15}\Omega
M^{-1}$: 
\begin{align}
\parbox{140mm}{\begin{picture}(120,35)
\put(-20,-153){\epsfig{file=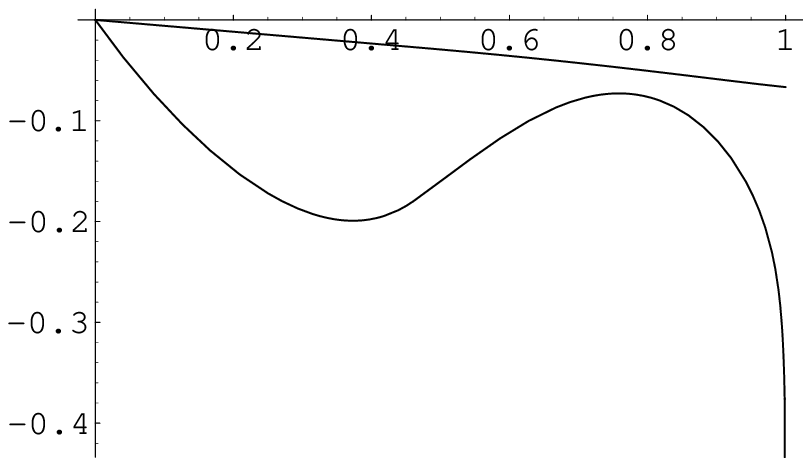,scale=0.7,bb=0 0 598 843}}
\put(70,20){\begin{tabular}{p{60mm}}
Comparison of \mbox{$E(\Omega,M^{-1})$} (the lower curve)
with \mbox{$-\frac{1}{15}\Omega M^{-1}$} (the upper curve),
both plotted over \mbox{$\Omega$}
\end{tabular}}
\end{picture}}  
\end{align}
This finishes the proof. \hfill $\square$
\bigskip

For convenience we give a plot of the scale function $M$:
\begin{align}
\parbox{120mm}{\begin{picture}(120,35)
\put(-20,-153){\epsfig{file=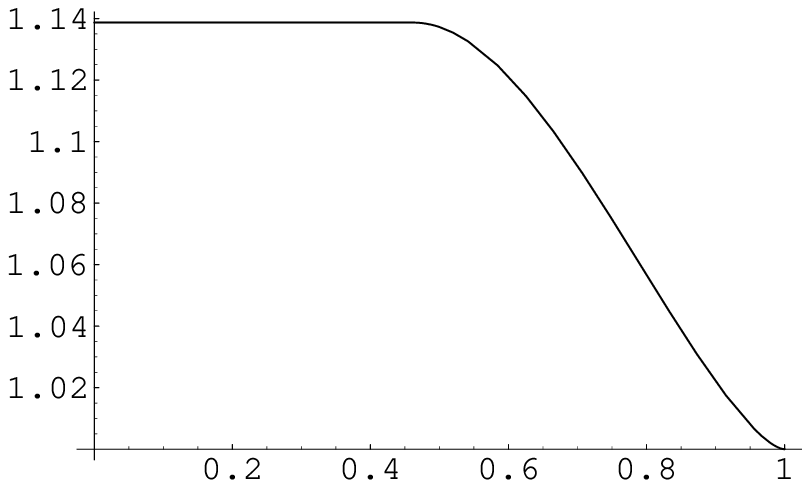,scale=0.7,bb=0 0 598 843}}
\put(70,20){\begin{tabular}{l}
The scale function \mbox{$M$} \\plotted over \mbox{$\Omega$}
\end{tabular}}
\end{picture}}  
\end{align}

\subsection{Main scaled bounds}

The first result is to prove that the propagator shows a scaled 
exponential decay in any index. This is expressed by 
\begin{theorem}
\label{thm-th1}
There exists a constant $K$ such that for $\Omega\in [0.5,1)$, we have the 
uniform  bound
  \begin{equation} 
    \label{th1}
    G^i_{m, m+h; l + h, l }\les%
    KM^{-i} e^{-\frac{\Omega}{15}M^{-i} \|m+l+h\|}\;,    
  \end{equation}
where the scale parameter $M(\Omega)>1$ is given by (\ref{M-1}) and
(\ref{Rbeta}). 
\end{theorem}
{\bf Proof.} For the first slice $i=1$ we use the bound (\ref{inijump-2}). With
(\ref{eq:exp1}) and $\sqrt{ml}\les \frac{1}{2}(m+l)$ we obtain 
\begin{align}
G^{(\alpha)}_{m,m+h;l+h,l} \les 
\exp\Big( (m+l+h) E(\alpha,\Omega)  \Big) \;.
\end{align}
Now, the bound (\ref{th1}) for the first slice $i=1$ follows from 
Lemma~\ref{Lemma1}, provided that $M^{-1}$ is chosen according to
(\ref{M-1}) and (\ref{Rbeta}).  

Next, for $i\ges 2$, we use the bound (\ref{inijump}), which with
(\ref{eq:exp1}) can be brought into the form
\begin{align}
G^{(\alpha)}_{m,m+h;l+h,l} 
&\les 
\exp\left(\alpha(m+l+h) \left(
\frac{(1-\Omega^2)}{4\Omega\sqrt{1-\alpha}}
+ \frac{1}{\alpha} \ln  \frac{4\Omega\sqrt{1-\alpha}}{4\Omega 
+ (1-\Omega)^2\alpha} \right)\right)\;.
\label{th1-exp}
\end{align}
We have to prove that the exponent 
\begin{align}
\hat{E}_\beta(\Omega,\alpha)
:= \beta \frac{(1-\Omega^2)\alpha}{4\Omega\sqrt{1-\alpha}}
+  \ln  \frac{4\Omega\sqrt{1-\alpha}}{4\Omega + (1-\Omega)^2\alpha}
\label{hatE}
\end{align}
is negative for $\beta=1$ and all $\alpha \les M^{-1}$,
where $M^{-1}$ is determined by the first slice. One has 
$\hat{E}_\beta(\Omega,0)=0$ and
\begin{align}
\frac{\partial}{\partial \alpha} 
\hat{E}_\beta(\Omega,\alpha) 
&= \beta\frac{(1-\Omega^2)(2-\alpha)}{8\Omega\sqrt{(1-\alpha)^3}}
-\frac{2(1+\Omega^2) -\alpha(1-\Omega)^2}{
2(1-\alpha)(4\Omega+(1-\Omega)^2\alpha)}\;.
\end{align}
This implies $\frac{\partial}{\partial \alpha}
\hat{E}_\beta(\Omega,\alpha)|_{\alpha=0}  =
\frac{(\beta-1)(1-\Omega^2)}{4\Omega}- \frac{\Omega}{2}$ and
$\frac{\partial}{\partial \alpha} \hat{E}_\beta(\Omega,\alpha)
|_{\alpha=1}=+\infty$.  Thus, for
$\Omega$ large enough and $\alpha$ small enough, we have $\frac{1}{\alpha}
\hat{E}_\beta(\Omega,\alpha)<0$. As $\alpha$ increases,
$\hat{E}_\beta(\Omega,\alpha)$ will remain negative up to some $\alpha_\beta$
with $\hat{E}_\beta(\Omega,\alpha_\beta)=0$. The next plot shows
$\frac{1}{\alpha}\hat{E}_1(\Omega, \alpha)$ for certain values of $\Omega$:
\begin{align}
\parbox{140mm}{\begin{picture}(120,35)
\put(-20,-153){\epsfig{file=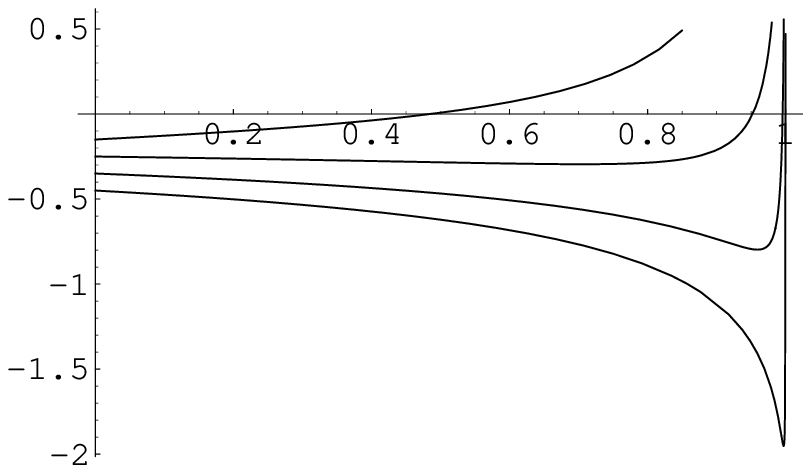,scale=0.7,bb=0 0 598 843}}
\put(35,32){\mbox{\scriptsize$\Omega=0.3$}}
\put(44,10){\mbox{\scriptsize$\Omega=0.9$}}
\put(70,20){\begin{tabular}{p{60mm}}
The function \mbox{$\frac{1}{\alpha}E_1(\Omega,\alpha)$}, for 
\mbox{$\Omega\in\{0.3,\;0.5,\;0.7,\;0.9\}$} plotted over 
\mbox{$\alpha$}. The larger the value of \mbox{$\Omega$}, the larger is the
zero of \mbox{$\frac{1}{\alpha}E_1(\Omega,\alpha)$}.
\end{tabular}}
\end{picture}}  
\end{align}
We have to ensure that $\alpha_1 > M^{-1}$. Thus, if 
$\hat{E}_\beta(\Omega,M^{-1})<0$, which requires
an $\Omega$ large enough, there will exist a constant $c>0$ such that
$\frac{1}{\alpha} \hat{E}_\beta(\Omega,\alpha) \les -c$ for all $\alpha
\in[0,M^{-1}]$. The critical value for $\Omega$ is found when plotting
$M \hat{E}_\beta(\Omega,M^{-1})$ over $\Omega$. Comparing
$M \hat{E}_1(\Omega,M^{-1})$ with the curve $-\frac{1}{2}\Omega$ relevant for
$\alpha=0$, 
\begin{align}
\parbox{140mm}{\begin{picture}(140,35)
\put(-20,-153){\epsfig{file=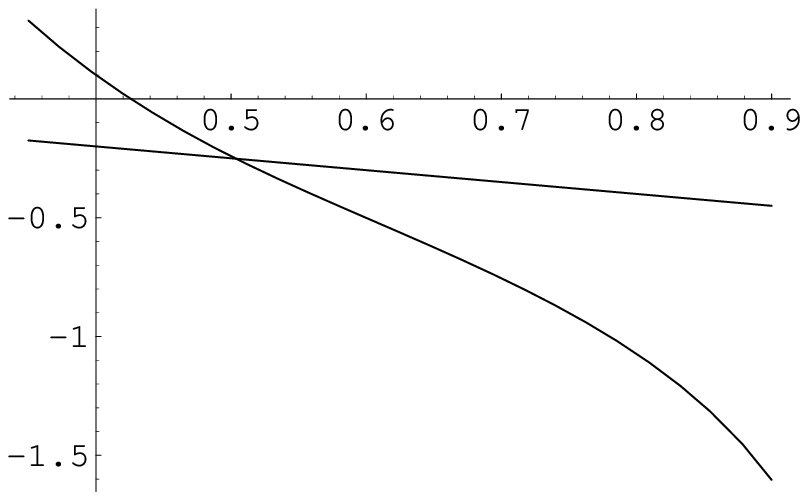,scale=0.7,bb=0 0 598 843}}
\put(70,18){\begin{tabular}{p{60mm}}
Comparison of \mbox{$M E_1(\Omega,M^{-1})$} (the lower curve at large
\mbox{$\Omega$} ) with 
\mbox{$-\frac{1}{2}\Omega$}, 
both plotted over \mbox{$\Omega$}. 
\end{tabular}}
\end{picture}}  
\label{omega-11}
\end{align}
we see that for $\Omega\ges 0.5$, the following estimation holds:
\begin{align}
G^{(\alpha)}_{m,m+h,l+h,h} \les e^{-\frac{1}{15}\Omega (m+l+h)
\alpha}\;.   
\end{align}
The Theorem now follows from (\ref{prop-slice-i}), with
$K=\frac{\theta(M_1-1)}{8\Omega}$. \hfill $\square$
\bigskip

\begin{theorem}
\label{thm-th2}
  For the scale parameter $M$ according to (\ref{M-1}) and (\ref{Rbeta}) 
there
exists a constant\footnote{In the following, the $K$'s will be kinds of
``dustbin'' constants. It means that their contents changes whereas their
names do not.} $K$ such that for all $\Omega \in [0.5,1)$ we 
have the uniform bound
\begin{align} 
&   G^i_{m, m+h; l + h, l }
\nonumber
\\*
& \les \frac{K}{M^{i}} 
e^{-\frac{\Omega}{15}M^{-i} \|m+l+h\|} 
\prod_{s=1}^{\frac{D}{2}} \min\lbt 1, 
\lbt
    \frac{1+\min(m^s,l^s,m^s+h^s,l^s+h^s)}{M^i/5}
\rbt^{\!\!\frac{|m^s-l^s|}{2}} \rbt.
\label{th2}
\end{align}
\end{theorem}
{\bf Proof.} Of course, this bound improves (\ref{th1}) only when an 
index component is smaller than $M^{i}/5$. We can, therefore, assume that 
$i> 12$. In particular, there is nothing to prove for the first slice
$i=1$. 

Suppose $l\les m\les m+h$ and $\delta=m-l$. Instead of (\ref{eq:exp1}) we use
the improved estimation
\begin{align}
  \sum_{u=0}^l \frac{X^{m-u}}{(m-u)!} \frac{Y^{l-u}}{(l-u)!} 
&= \sum_{v=0}^l \frac{X^{m-l+v}}{(m-l+v)!} \frac{Y^v}{v!}
\nonumber
\\
&\les  \frac{X^{m-l}}{(m-l)!} \sum_{v=0}^l \left ( \frac{1}{v!} \right )^2
(XY)^v
\les  \frac{X^{m-l}}{(m-l)!} e^{X+Y}\;.
\label{ineq2}
\end{align}
Then, the propagator (\ref{inijump}) takes the form
\begin{align}
G^{(\alpha)}_{m,m+h;l+h,l} 
&\les 
\exp\left((m+l+h)\alpha \left( \frac{(1-\Omega^2)}{4\Omega \sqrt{1-\alpha}} 
+ \frac{1}{\alpha} \ln \frac{4\Omega\sqrt{1-\alpha}}{
4\Omega + (1-\Omega)^2\alpha}\right)\right)
\nonumber
\\*
&  \times 
\sqrt{ \frac{\left(\dfrac{m\alpha (1-\Omega^2)}{
4\Omega \sqrt{1-\alpha}(2\beta-2)}\right)^\delta }{\delta!}}
\sqrt{ \frac{\left(\dfrac{(2\beta-2)(m+h)\alpha (1-\Omega^2)}{
4\Omega \sqrt{1-\alpha}}\right)^\delta }{\delta!}}\;.
\label{Delta-c}
\end{align}
We choose $\beta=\frac{5}{4}$, but any choice $\beta>1$ would be possible. In
the last term (containing $m+h$) we estimate $\frac{x^\delta}{\delta!} \les
e^x$ and add the exponential to the first line of (\ref{Delta-c}). In the
first term of the last line of (\ref{Delta-c}) we use the estimation
\begin{equation}
\frac{1}{\delta!} \les \lbt\frac{\delta}{e}\rbt^{-\delta}\;.
\label{stirling}
\end{equation}
This yields
\begin{align}
&G^{(\alpha)}_{m,m+h;l+h,l} 
\nonumber
\\*
&\les \left(\dfrac{m\alpha (1-\Omega^2)e}{
2\Omega \delta \sqrt{1-\alpha}}\right)^{\frac{\delta}{2}}\;
\exp\left((m{+}l{+}h)\alpha \left( \frac{5}{4} \frac{(1-\Omega^2)}{
4\Omega \sqrt{1-\alpha}} 
+ \frac{1}{\alpha} \ln \frac{4\Omega\sqrt{1-\alpha}}{
4\Omega + (1-\Omega)^2\alpha}\right)\right)\;.
\label{Delta-c-2}
\end{align}
For $\beta=\frac{5}{4}$ the exponent in (\ref{Delta-c-2}) is negative 
for $\Omega\ges 0.5$:
\begin{align}
\parbox{140mm}{\begin{picture}(140,35)
\put(-20,-153){\epsfig{file=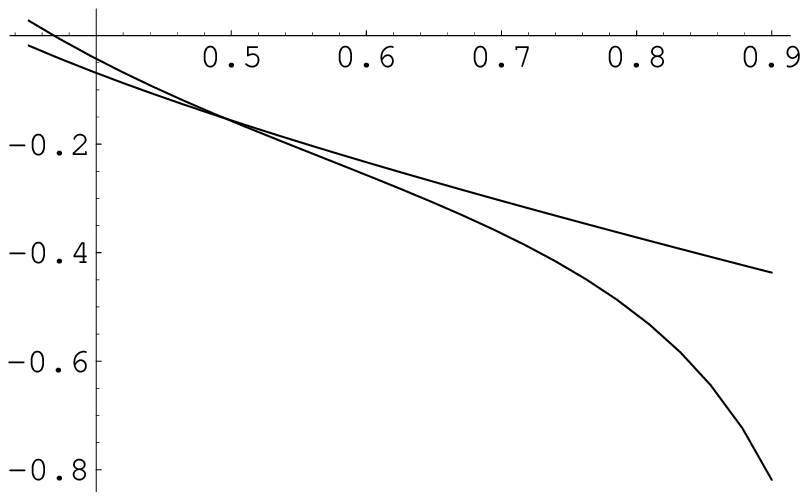,scale=0.7,bb=0 0 598 843}}
\put(70,18){\begin{tabular}{p{60mm}}
Comparison of \mbox{$M^{12} E_{\frac{5}{4}}(\Omega,M^{-12})$} 
(the lower curve at large \mbox{$\Omega$}) with 
\mbox{$\frac{1-\Omega^2}{16\Omega} -\frac{1}{2}\Omega$}, 
both plotted over \mbox{$\Omega$}. 
\end{tabular}}
\end{picture}}  
\end{align}
Next, for $i\ges 12$ and $\Omega \ges 0.5$, one has
$\frac{ (1-\Omega^2)e}{\Omega \sqrt{1-M^{-i}}} \les 5$, as the following plot 
shows:
\begin{align}
\parbox{140mm}{\begin{picture}(120,35)
\put(-20,-153){\epsfig{file=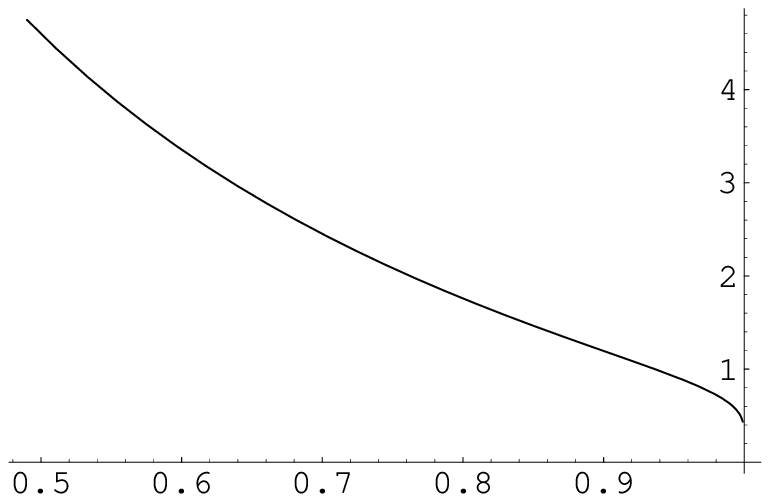,scale=0.7,bb=0 0 598 843}}
\put(70,18){\begin{tabular}{p{45mm}}
The function \mbox{$\frac{ (1-\Omega^2)e}{\Omega \sqrt{1-M^{-12}}}$}
plotted over \mbox{$\Omega$}. 
\end{tabular}}
\end{picture}}  
\end{align}
Now we are left with two cases:
  \begin{itemize}
  \item[a)] $l=0 \Leftrightarrow m=\delta$:
    \begin{equation}
      G^{(\alpha)}_{m, m+h; l + h, l }\les 
      e^{-\frac{\Omega}{15} (m+l+h) \alpha} 
\left(\tfrac{5}{2}\alpha\right)^{\delta/2} \;.
    \end{equation}
  \item[b)] $l\ges 1$ and $\delta\ges 1$: using $l+\delta\les 2l\delta$,
    \begin{equation} \label{lowindex}
      G^{(\alpha)}_{m, m+h; l + h, l }\les 
      e^{-\frac{\Omega}{15}(m+l+h) \alpha} 
 \left( 5 l \alpha\right)^{\delta/2} \;.
    \end{equation}
  \end{itemize}
Inserting this into (\ref{prop-slice-i}) and symmetrizing with
respect to the smallest index we obtain (\ref{th2}) with
$K=\frac{\theta(M-1)}{8\Omega}$.  \hfill$\square$

\bigskip

Let us now consider a typical graph appearing in the process of
renormalization, that is, with external legs carrying indices
\emph{lower} than the internal ones. The bound (\ref{th2}) provides a
good factor with respect to power-counting unless the index jump
$\delta=|m-l|$ is very small, typically $\delta=0$ or $\delta=1$. This
ensures that if the lower index of a propagator is smaller than the
scale we look at, the index is conserved along its trajectory 
for power-counting relevant and marginal graphs.

Unfortunately, that estimation does not carry any information when the lower
index is larger than the scale. It leads to a difficulty for graphs
which possess completely inner vertices. Therefore, we have to find estimates 
for propagators with a sum over the index $l$. The next section is
devoted to these bounds.


\subsection{Bounds for sums}

Now we want to prove that the summation of the propagator
$G^{(\alpha)}_{m,p-l,p,m+l}$ over $l$, for $m$ and $p$ 
kept constant, gives the same power-counting as in the previous section. The
proof relies on a more accurate estimate of the sum in (\ref{eq:exp1}).
\begin{theorem}
For $\Omega \in [0.5,1)$ there exists a constant $K$ such
that we have the uniform bound
\begin{align}
\sum_{l =-m}^p G^i_{m,p-l,p,m+l} &\les 
K M^{-i} \,e^{-\frac{\Omega}{20} M^{-i} (\|p\|+\|m\|) }\;.
\label{thsum}
\end{align}
\end{theorem}
{\bf Proof.} The first slices, say $i \les 16$, are trivial to treat. Using
(\ref{th1}) we have
\begin{align}
\sum_{l =-m}^p G^i_{m,p-l,p,m+l} \les \frac{K}{M^i} \prod_{s=1}^2 
(m^s+p^s+1) e^{-\frac{\Omega M^{-i}}{15}(m^s+p^s)}\;.
\label{xexpx-1}
\end{align}
Then, the estimation follows from
\begin{align}
(x+1) e^{-\frac{1}{15} \Omega M^{-i} x } \les \left(\frac{60 M^i}{\Omega} 
e^{\frac{\Omega M^{-i}}{60}-1}\right) 
e^{- \frac{1}{20} \Omega M^{-i} x}\;.
\label{xexpx-2}
\end{align}

This method fails in the limit $i\to \infty$. Thus, for large $i$, we have to
estimate the propagator (\ref{inijump}) more carefully, now putting $h\mapsto
p-m-l$ and $l\mapsto m+l$.  Without loss of generality we can assume $p \ges
m$. We have to divide the range of summation into three parts according to the
smallest index. Using (\ref{ineq2}) we estimate
\begin{align}
&\sum_{l=-m}^p G^{(\alpha)}_{m, p-l; p, m+l}
\nonumber
\\*
&\les \left(\frac{4\Omega\sqrt{1-\alpha}}{4\Omega+(1-\Omega)^2\alpha} 
\right)^{m+p}  \Bigg(
\sum_{l=-m}^{-1} \sum_{u=0}^{m+l}
\frac{X^{m-u}}{(m-u)!} \frac{ Y^{m+l-u}}{(m+l-u)!}
\nonumber
\\*
&+\sum_{l=0}^{p-m-1} \sum_{u=0}^{m}
\frac{X^{m-u}}{(m-u)!} \frac{ Y^{m+l-u}}{(m+l-u)!}
+\sum_{l=p-m}^{p} \sum_{v=0}^{p-l}
\frac{X^{p-l-v}}{(p-l-v)!} \frac{ Y^{p-v}}{(p-v)!} \Bigg)
\nonumber
\\*
& \les \left( 
\sum_{l=-m}^{-1}\frac{X^{|l|}}{|l|!}
+\sum_{l=0}^{p}\frac{Y^{l}}{l!}\right)\;
\exp\left((m+p)\left(\frac{\alpha (1-\Omega^2)}{
4\Omega\sqrt{1-\alpha}}
+ \ln \frac{4\Omega\sqrt{1-\alpha}}{4\Omega+(1-\Omega)^2\alpha}\right)\right)
\nonumber
\\*
&\les 
 2 \underbrace{\sum_{l=0}^p \frac{
\left(\frac{\alpha (1-\Omega^2)(m{+}p{+}l)}{
8\Omega\sqrt{1-\alpha}}\right)^l}{l!}}_{Z}
 \exp\left((m+p)\alpha\left(\frac{(1-\Omega^2)}{4\Omega\sqrt{1-\alpha}} 
+ \frac{1}{\alpha} 
\ln \frac{4\Omega\sqrt{1-\alpha}}{4\Omega+(1-\Omega)^2\alpha}\right)
\right)\;,
\raisetag{2ex}
\label{Deltaq}
\end{align}
where $X=\frac{\alpha (1-\Omega^2)(p+m-l)}{8\Omega\sqrt{1-\alpha}}$ and 
$Y=\frac{\alpha (1-\Omega^2)(p+m+l)}{8\Omega\sqrt{1-\alpha}}$.

We can now divide the sum over $l$
into two regions corresponding to $l\les (2\beta-3) (p+m)$ and $l \ges
\left\lceil(2\beta-3)(p+m)\right\rceil\ges (2\beta-3)(p+m)$, where $\left\lceil
  x\right\rceil$ is the smallest integer which is larger than $x$ and
$\beta>\frac{3}{2}$ will be determined later:
\begin{align}
Z &\les \sum_{l = 0}^{(2\beta-3)(p+m)}
 \frac{\left(
\frac{\alpha (1-\Omega^2)(\beta-1)(m+p)}{4\Omega\sqrt{1-\alpha}}
\right)^l}{l!} 
+ \sum_{l=\left\lceil(2\beta-3)(p+m)\right\rceil}^p 
 \frac{\left(
\frac{\alpha (1-\Omega^2)(\beta-1)l}{4\Omega\sqrt{1-\alpha}(2\beta-3)}
\right)^l}{l!} \;.
\label{Z-0}
\end{align}
We extend both sums to infinity and use in 
the second one the identity (\ref{stirling}):
\begin{align}
Z &\les\exp \left(
\frac{\alpha (1-\Omega^2)(\beta-1)(m+p)}{4\Omega\sqrt{1-\alpha}}\right)
+ \sum_{l = 0}^\infty 
\left(\frac{\alpha (1-\Omega^2) e (\beta-1)}{4\Omega  \sqrt{1-\alpha}
(2\beta-3)}\right)^l\;.
\label{Z1}
\end{align}
For $\beta=\frac{39}{25}$ one confirms 
$\frac{M^{-16} (1-\Omega^2) e (\beta-1)}{4\Omega  \sqrt{1-M^{-16}}
(2\beta-3)} < 1$ for $\Omega \ges 0.5$: 
\begin{align}
\parbox{140mm}{\begin{picture}(140,35)
\put(-20,-153){\epsfig{file=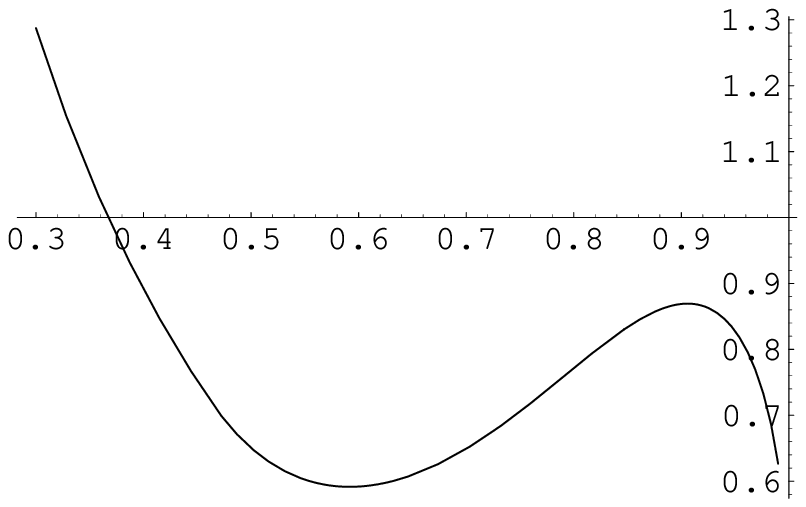,scale=0.7,bb=0 0 598 843}}
\put(70,18){\begin{tabular}{p{60mm}}
The function \mbox{$\frac{M^{-16} (1-\Omega^2) e (\beta-1)}{
4\Omega  \sqrt{1-M^{-16}}(2\beta-3)}$} for \mbox{$\beta=\frac{39}{25}$} 
plotted over \mbox{$\Omega$}. 
\end{tabular}}
\end{picture}}  
\end{align}
On the other hand, the following plot shows that
$\frac{39}{25} \frac{(1-\Omega^2)}{4\Omega\sqrt{1-\alpha}} 
+ \frac{1}{\alpha} 
\ln \frac{4\Omega\sqrt{1-\alpha}}{4\Omega+(1-\Omega)^2\alpha}
< -\frac{1}{20} \Omega$ for $\alpha\les M^{-16}$ and  
$\Omega \ges 0.5$:
\begin{align}
\parbox{140mm}{\begin{picture}(140,35)
\put(-20,-153){\epsfig{file=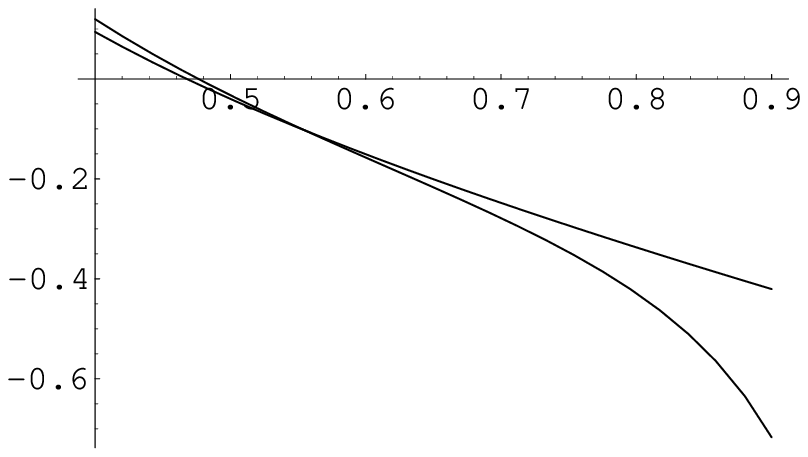,scale=0.7,bb=0 0 598 843}}
\put(70,18){\begin{tabular}{p{60mm}}
Comparison of \mbox{$M^{16} E_{\frac{39}{25}}(\Omega,M^{-16})$} 
(the lower curve at large \mbox{$\Omega$}) with 
\mbox{$\frac{7(1-\Omega^2)}{50\Omega}-\frac{1}{2}\Omega$}, 
both plotted over \mbox{$\Omega$}. 
\end{tabular}}
\end{picture}}  
\end{align}
This finishes the proof. \hfill $\square$
\bigskip

The previous estimation for the summed propagator is still not enough for the
renormalization proof, because the index sums are entangled in the graph. We
have to prove that the exponential decay is still achieved if for given
summation variable $l$ we maximise the other index $p$:
\begin{theorem}
  For $\Omega \ges 0.58$ there exists a constant $K$
  such that we have the uniform bound
  \begin{equation} \label{thsummax}
   \sum_{l=-m}^\infty \max_{p \ges \max(l,0)}G^i_{m,p-l;p,m+l}
\les  KM^{-i} e^{-\frac{\Omega}{36} M^{-i}\|m\|} \;.   
  \end{equation}
\end{theorem}
{\bf Proof.} Again, the main estimate (\ref{th1}) guarantees the desired
behavior (\ref{thsummax}) for the first slices, say $i \les 16$. For $l\les
0$ the maximum is attained at $p=0$ so that we are in the situation of
(\ref{xexpx-1}) and (\ref{xexpx-2}). For $l>0$ the maximum is attained at
$p=l$ so that the $l$-sum leads to a geometric series. Here, it is important
that $i$ is bounded.

For $i>16$ we have to proceed differently. We divide the domain of summation
according to the smallest index at the propagator:
\begin{align}
\sum_{l=-m}^\infty \max_{p \ges \max(l,0)}
= \sum_{l=-m}^{-1} \max_{0 \les p < m+l}
+ \sum_{l=-m}^{-1} \max_{m+l \les p < m}
+ \sum_{l=-m}^{-1} \max_{p \ges m}
+ \sum_{l=0}^{\infty} \max_{l \les p < m+l}
+ \sum_{l=0}^{\infty} \max_{p \ges m+l}\;.
\end{align}
We now obtain from (\ref{inijump}) and (\ref{ineq2})
\begin{align}
&\sum_{l=-m}^\infty \max_{p \ges \max(l,0)}G^{(\alpha)}_{m, p-l;
  p, m+l }
\nonumber
\\*
& \les \sum_{l=-m}^{-1} \max_{0 \les p < m+l} 
\left(\frac{4\Omega\sqrt{1-\alpha}}{4\Omega+(1-\Omega)^2\alpha} 
\right)^{m+p} 
\sum_{v=0}^p \frac{X^{p-l-v}}{(p-l-v)!}
\frac{Y^{p-v}}{(p-v)!}
\nonumber
\\*
& + \sum_{l=-m}^{-1} \max_{m+l \les p < m} 
\left(\frac{4\Omega\sqrt{1-\alpha}}{4\Omega+(1-\Omega)^2\alpha} 
\right)^{m+p} 
\sum_{u=0}^{m+l} \frac{X^{m-u}}{(m-u)!}
\frac{Y^{m+l-u}}{(m+l-u)!}
\nonumber
\\*
& + \sum_{l=-m}^{-1} \max_{p \ges m} 
\left(\frac{4\Omega\sqrt{1-\alpha}}{4\Omega+(1-\Omega)^2\alpha} 
\right)^{m+p} 
\sum_{u=0}^{m+l} \frac{X^{m-u}}{(m-u)!}
\frac{Y^{m+l-u}}{(m+l-u)!}
\nonumber
\\*
& + \sum_{l=0}^\infty \max_{l \les p < m+l} 
\left(\frac{4\Omega\sqrt{1-\alpha}}{4\Omega+(1-\Omega)^2\alpha} 
\right)^{m+p} 
\sum_{v=0}^{p-l} \frac{X^{p-l-v}}{(p-l-v)!}
\frac{Y^{p-v}}{(p-v)!}
\nonumber
\\*
& + \sum_{l=0}^\infty \max_{p \ges m+l} 
\left(\frac{4\Omega\sqrt{1-\alpha}}{4\Omega+(1-\Omega)^2\alpha} 
\right)^{m+p}  \sum_{u=0}^{m} \frac{X^{m-u}}{(m-u)!}
\frac{Y^{m+l-u}}{(m+l-u)!}
\nonumber
\\*
& \les \sum_{|l|=1}^{m} \max_{p \ges 0} 
\exp\left((p+m)\alpha\left(
\frac{(1-\Omega^2)}{4\Omega\sqrt{1-\alpha}}+\frac{1}{\alpha}
\ln \frac{4\Omega\sqrt{1-\alpha}}{4\Omega+(1-\Omega)^2\alpha} \right)\right)
\frac{\left(\frac{\alpha(1-\Omega^2)(p+m+|l|)}{
8\Omega\sqrt{1-\alpha}}\right)^{|l|}}{|l|!} 
\label{pmax1}
\\*
& + \sum_{l=0}^{\infty} \max_{p \ges l}
\exp\left((p+m)\alpha\left(
\frac{(1-\Omega^2)}{4\Omega\sqrt{1-\alpha}}+\frac{1}{\alpha}
\ln \frac{4\Omega\sqrt{1-\alpha}}{4\Omega+(1-\Omega)^2\alpha} \right)\right)
\frac{\left(\frac{\alpha(1-\Omega^2)(p+m+l)}{
8\Omega\sqrt{1-\alpha}}\right)^{l}}{l!} \;,
\label{pmax2}
\end{align}
where $X=\frac{\alpha (1-\Omega^2)(p+m-l)}{8\Omega\sqrt{1-\alpha}}$ and 
$Y=\frac{\alpha (1-\Omega^2)(p+m+l)}{8\Omega\sqrt{1-\alpha}}$.

The function $e^{-\gamma p} \frac{(p+q)^l}{l!}$ attains its maximum 
$e^{\gamma q-l} \,\frac{1}{l!}\left(\frac{l}{\gamma}\right)^l$ at  
$p=\frac{l}{\gamma}-q$. We need this property for $q=m+|l|$ and 
\begin{align}
\gamma:=-
\frac{\alpha(1-\Omega^2)}{4\Omega\sqrt{1-\alpha}}-
\ln \frac{4\Omega\sqrt{1-\alpha}}{4\Omega+(1-\Omega)^2\alpha} >0\;.
\label{gamma}
\end{align} 
However, we have to take into account the range of
$p$. If $\frac{l}{\gamma}-q<0$ in (\ref{pmax1}), then the function 
$e^{-\gamma p} \frac{(p+q)^l}{l!}$ is monotonously decreasing for all $p\ges
0$ so that the maximum is at $p=0$. Otherwise we have to insert the specific
maximum. This means that we split the sum over $l$ in (\ref{pmax1}) 
into two pieces 
\begin{itemize}
\item $|l|\les \frac{m\gamma}{1-\gamma}$, where we insert $p=0$. Actually, we
  can safely extend this sum from $0$ to $m$, still keeping $p=0$.

\item $|l|\ges \frac{m\gamma}{1-\gamma}$, where we insert
  $p=\frac{1-\gamma}{\gamma} |l|-m$. 
\end{itemize}
This gives 
\begin{align}
& \sum_{|l|=1}^{m} \max_{p \ges 0} 
\exp\left((p+m)\alpha\left(
\frac{(1-\Omega^2)}{4\Omega\sqrt{1-\alpha}}+\frac{1}{\alpha}
\ln \frac{4\Omega\sqrt{1-\alpha}}{4\Omega+(1-\Omega)^2\alpha} \right)\right)
\frac{\left(\frac{\alpha(1-\Omega^2)(p+m+|l|)}{
8\Omega\sqrt{1-\alpha}}\right)^{|l|}}{|l|!}
\nonumber
\\*
& \les \sum_{l=0}^m 
\exp\left(m\alpha\left(
\frac{(1-\Omega^2)}{4\Omega\sqrt{1-\alpha}}+\frac{1}{\alpha}
\ln \frac{4\Omega\sqrt{1-\alpha}}{4\Omega+(1-\Omega)^2\alpha} \right)\right)
\frac{\left(\frac{\alpha(1-\Omega^2)(m+l)}{
8\Omega\sqrt{1-\alpha}}\right)^{l}}{l!}
\nonumber
\\*
& + \sum_{l=\frac{\gamma m}{1-\gamma}}^m 
\exp\left(-l-\alpha l\left(
\frac{(1-\Omega^2)}{4\Omega\sqrt{1-\alpha}}+\frac{1}{\alpha}
\ln \frac{4\Omega\sqrt{1-\alpha}}{4\Omega+(1-\Omega)^2\alpha} \right)\right)
\frac{\left(\frac{\alpha(1-\Omega^2)l}{
8\Omega\gamma \sqrt{1-\alpha}}\right)^{l}}{l!}\;.
\end{align} 

For the first sum we are in the situation of (\ref{Deltaq}) with $m\mapsto
0,p\mapsto m$ so that we can bound that sum according to the steps leading to
(\ref{thsum}) by a constant times $e^{-\frac{1}{20}\Omega m \alpha}$, for
$\Omega \ges 0.5$ and $\alpha \les M^{-16}$.
In the second sum we use (\ref{stirling}) 
so that, for some number $\epsilon>0$,
\begin{align}
&\sum_{l=\frac{\gamma m}{1-\gamma}}^m 
\exp\left(-l-\alpha l\left(
\frac{(1-\Omega^2)}{4\Omega\sqrt{1-\alpha}}+\frac{1}{\alpha}
\ln \frac{4\Omega\sqrt{1-\alpha}}{4\Omega+(1-\Omega)^2\alpha} \right)\right)
\frac{\left(\frac{\alpha(1-\Omega^2)l)}{
8\Omega\gamma \sqrt{1-\alpha}}\right)^{l}}{l!}
\nonumber
\\*
& \les 
\sum_{l=\frac{\gamma m}{1-\gamma}}^m e^{-\epsilon l}
\left(\frac{\alpha(1-\Omega^2)e^{\epsilon+\gamma}}{
8\Omega\gamma \sqrt{1-\alpha}}\right)^{l}
\les 
e^{-\epsilon \gamma m} \sum_{l=\frac{\gamma m}{1-\gamma}}^m 
\left(\frac{\alpha(1-\Omega^2)e^{\epsilon+\gamma}}{
8\Omega\gamma \sqrt{1-\alpha}}\right)^{l}\;.
\end{align}
In the numerator we can bound $\gamma$ by $\frac{1}{2}$, otherwise the sum
vanishes. Moreover, we choose $\epsilon=\frac{1}{16}$. We then see from
(\ref{gamma}) that the following condition is to prove: 
\begin{align}
\frac{8\Omega\gamma \sqrt{1-\alpha}}{\alpha(1-\Omega^2)e^{\frac{9}{16}}}
= \frac{8\Omega \sqrt{1-\alpha}}{\alpha(1-\Omega^2)e^{\frac{9}{16}}}
\ln \frac{4\Omega+(1-\Omega)^2\alpha}{4\Omega\sqrt{1-\alpha}}
- \frac{2}{e^{\frac{9}{16}}}
> 1 \;.
\end{align}
This function is monotonously decreasing in $\alpha$ so that it is sufficient
to check it for  $\alpha=M^{-16}$:
\begin{align}
\parbox{140mm}{\begin{picture}(140,35)
\put(-20,-153){\epsfig{file=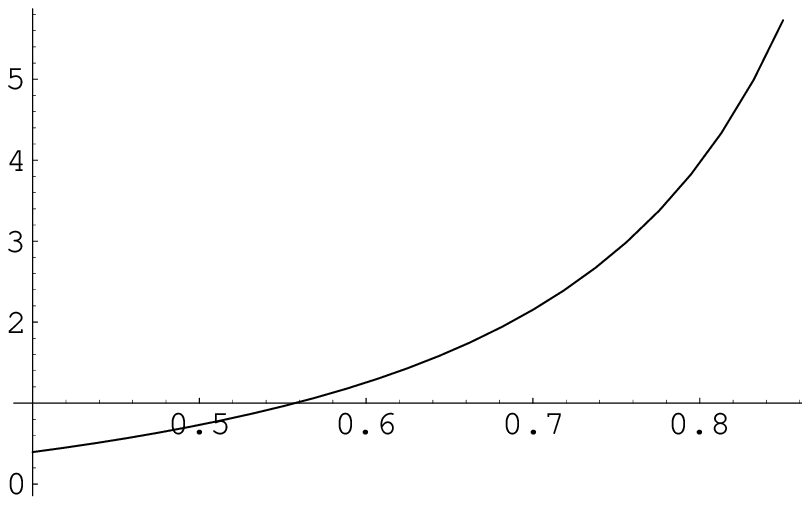,scale=0.7,bb=0 0 598 843}}
\put(70,18){\begin{tabular}{l}
The function \\ \mbox{$ \frac{8\Omega \sqrt{1-M^{-16}}}{
M^{-16}(1-\Omega^2)e^{\frac{9}{16}}}
\ln \frac{4\Omega+(1-\Omega)^2M^{-16}}{4\Omega\sqrt{1-M^{-16}}}
- \frac{2}{e^{\frac{9}{16}}}$} \\ plotted over \mbox{$\Omega$}. 
\end{tabular}}
\end{picture}}  
\end{align}
We see that it is sufficient to have $\Omega\ges 0.58$. Moreover, 
$-\epsilon \gamma m= \frac{\alpha}{16} m (\frac{1}{\alpha} 
\hat{E}_1(\Omega,\alpha))$, where
$\hat{E}_1$ is given in (\ref{hatE}). We then know from (\ref{omega-11}) that
for $\alpha \in [0,M^{-16}]$ and $\Omega \ges 0.58$,  
$-\epsilon \gamma m \les - \frac{1}{32} \alpha m$. 
This shows that (\ref{pmax1}) leads to the desired estimation
(\ref{thsummax}).

We now pass to (\ref{pmax2}). The condition $p\ges l$ leads to a splitting of
the $l$-sum at $\frac{\gamma m}{1-2\gamma}$. For smaller $l$ we have $p=l$:
\begin{align}
& \sum_{l=0}^{\infty} \max_{p \ges l} 
\exp\left((p+m)\alpha\left(
\frac{(1-\Omega^2)}{4\Omega\sqrt{1-\alpha}}+\frac{1}{\alpha}
\ln \frac{4\Omega\sqrt{1-\alpha}}{4\Omega+(1-\Omega)^2\alpha} \right)\right)
\frac{\left(\frac{\alpha(1-\Omega^2)(p+m+|l|)}{
8\Omega\sqrt{1-\alpha}}\right)^{l}}{l!}
\nonumber
\\*
& \les \sum_{l=0}^\infty 
\exp\left((m+l)\alpha\left(
\frac{(1-\Omega^2)}{4\Omega\sqrt{1-\alpha}}+\frac{1}{\alpha}
\ln \frac{4\Omega\sqrt{1-\alpha}}{4\Omega+(1-\Omega)^2\alpha} \right)\right)
\frac{\left(\frac{\alpha(1-\Omega^2)(m+2l)}{
8\Omega\sqrt{1-\alpha}}\right)^{l}}{l!}
\nonumber
\\*
& + \sum_{l=\frac{\gamma m}{1-2\gamma}}^\infty 
\exp\left(-l-\alpha l\left(
\frac{(1-\Omega^2)}{4\Omega\sqrt{1-\alpha}}+\frac{1}{\alpha}
\ln \frac{4\Omega\sqrt{1-\alpha}}{4\Omega+(1-\Omega)^2\alpha} \right)\right)
\frac{\left(\frac{\alpha(1-\Omega^2)l}{
8\Omega\gamma \sqrt{1-\alpha}}\right)^{l}}{l!}\;.
\label{pmax3}
\end{align} 
The second sum on the right hand side is identical to treat as in the previous case
(\ref{pmax1}).  The first sum on this right hand side 
is split as in (\ref{Z-0}) 
at $l=(\beta-\frac{3}{2})m$. Compared with (\ref{Z1}) we now have to achieve
$\frac{M^{-16} (1-\Omega^2) e (2\beta-2)}{4\Omega  \sqrt{1-M^{-16}}
(2\beta-3)} < 1$ for $\Omega \ges 0.58$. We can take $\beta=\frac{5}{3}$: 
\begin{align}
\parbox{140mm}{\begin{picture}(140,35)
\put(-20,-153){\epsfig{file=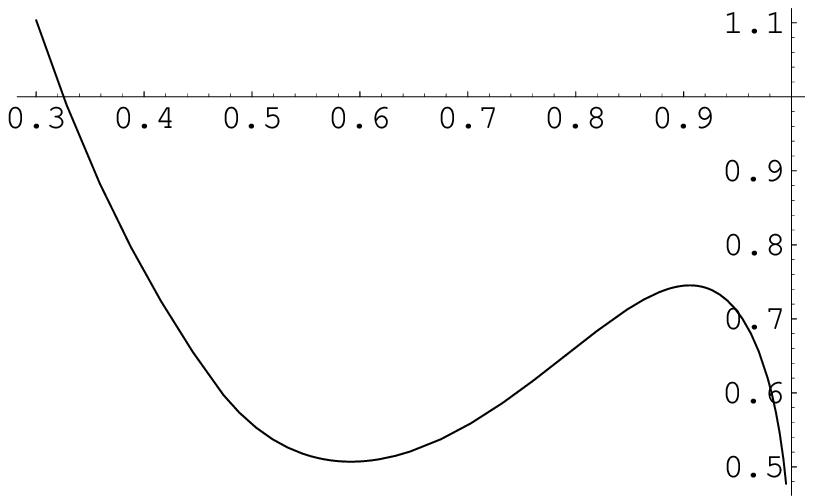,scale=0.7,bb=0 0 598 843}}
\put(70,18){\begin{tabular}{p{60mm}}
The function \mbox{$\frac{M^{-16} (1-\Omega^2) e (\beta-1)}{
4\Omega  \sqrt{1-M^{-16}}(2\beta-3)}$} for \mbox{$\beta=\frac{5}{3}$} 
plotted over 
\mbox{$\Omega$}. 
\end{tabular}}
\end{picture}}  
\end{align}
On the other hand, the following plot shows that
$\frac{5}{3} \frac{(1-\Omega^2)}{4\Omega\sqrt{1-\alpha}} 
+ \frac{1}{\alpha} 
\ln \frac{4\Omega\sqrt{1-\alpha}}{4\Omega+(1-\Omega)^2\alpha}
< -\frac{1}{36} \Omega$ for $\alpha\les M^{-16}$ and  
$\Omega \ges 0.58$:
\begin{align}
\parbox{140mm}{\begin{picture}(140,35)
\put(-20,-153){\epsfig{file=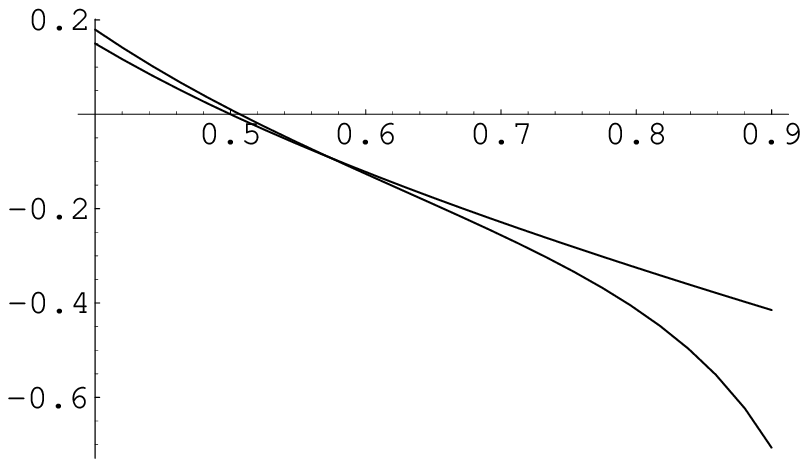,scale=0.7,bb=0 0 598 843}}
\put(70,18){\begin{tabular}{p{60mm}}
Comparison of \mbox{$M^{16} E_{\frac{5}{3}}(\Omega,M^{-16})$} 
(the lower curve at large \mbox{$\Omega$}) with 
\mbox{$\frac{(1-\Omega^2)}{6\Omega}-\frac{1}{2}\Omega$}, 
both plotted over \mbox{$\Omega$}. 
\end{tabular}}
\end{picture}}  
\end{align}
The $l$-dependence of the argument of the 
exponential in the second line of (\ref{pmax3}) can be ignored.  
This finishes the proof. \hfill $\square$
\bigskip

\subsection{Composite propagators}

This section is devoted to the proofs of bounds on the \emph{composite
  propagators} introduced in \cite{Grosse:2004yu}. We define their 
sliced versions as follows:
\begin{align}
\mathcal{Q}^{i(0)}_{\di{m^1}{m^2},
\di{n^1}{n^2};\di{n^1}{n^2},\di{m^1}{m^2}} 
&= G^i_{\di{m^1}{m^2},
\di{n^1}{n^2};\di{n^1}{n^2},\di{m^1}{m^2}} - G^i_{\di{0}{0},
\di{n^1}{n^2};\di{n^1}{n^2},\di{0}{0}}\;, 
\label{comp-0}
\\
\mathcal{Q}^{i(1)}_{\di{m^1}{m^2},
\di{n^1}{n^2};\di{n^1}{n^2},\di{m^1}{m^2}} 
&=\mathcal{Q}^{i(0)}_{\di{m^1}{m^2},
\di{n^1}{n^2};\di{n^1}{n^2},\di{m^1}{m^2}}
- m^1 \mathcal{Q}^{i(0)}_{\di{1}{0},
\di{n^1}{n^2};\di{n^1}{n^2},\di{1}{0}}
- m^2 \mathcal{Q}^{i(0)}_{\di{0}{1},
\di{n^1}{n^2};\di{n^1}{n^2},\di{0}{1}}\;,
\\
\mathcal{Q}^{i(\frac{1}{2})}_{\di{m^1+1}{m^2},
\di{n^1+1}{n^2};\di{n^1}{n^2},\di{m^1}{m^2}} 
&= G^i_{\di{m^1+1}{m^2},
\di{n^1+1}{n^2};\di{n^1}{n^2},\di{m^1}{m^2}} - \sqrt{m^1+1} 
G^i_{\di{1}{0},
\di{n^1+1}{n^2};\di{n^1}{n^2},\di{0}{0}}\;.
\end{align}

\begin{theorem} \label{thcomprop1}
For $M$ according to (\ref{M-1}) and (\ref{Rbeta}) there exist constants
$K_i$ such that for $\Omega\in [0.5,1)$ and $m\les M^{i}$, we
have the uniform bounds
\begin{align} 
\left|\mathcal{Q}^{i(0)}_{\di{m^1}{m^2},
\di{n^1}{n^2};\di{n^1}{n^2},\di{m^1}{m^2}}\right| 
&\les K_0 M^{-i}\, \frac{m^1+m^2}{M^{i}}\,
e^{- \frac{\Omega}{15} M^{-i} (n^1+n^2)}\;,
\label{thcomp0}
\\
\left|\mathcal{Q}^{i(1)}_{\di{m^1}{m^2},
\di{n^1}{n^2};\di{n^1}{n^2},\di{m^1}{m^2}}\right| & \les K_1 
M^{-i} \,\lbt\frac{m^1+m^2}{M^{i}}\rbt^{2}\,
e^{- \frac{\Omega}{15} M^{-i} (n^1+n^2)}\;,
\label{thcomp1}
\\
\left|\mathcal{Q}^{i(\frac{1}{2})}_{\di{m^1+1}{m^2},
\di{n^1+1}{n^2};\di{n^1}{n^2},\di{m^1}{m^2}}\right| &\les K_2 M^{-i} 
\lbt\frac{m^1+m^2+1}{M^{i}}\rbt^{3/2}\, 
e^{- \frac{\Omega}{15} M^{-i} (n^1+n^2)}\;.
\label{thcomp2}
\end{align}
\end{theorem}
\noindent
{\bf Proof.} 
There is no need to discuss the first slice. 
From (\ref{prop-slice-i}) we have
\begin{align}
\mathcal{Q}^{i(0)}_{\di{m^1}{m^2}, \di{n^1}{n^2}; 
\di{n^1}{n^2}, \di{m^1}{m^2}} 
= \frac{\theta}{8\Omega} \int_{M^{-i}}^{M^{-i+1}} d\alpha\,  
\dfrac{(1-\alpha)^{\frac{\mu_0^2 \theta}{8 \Omega}}}{  
(1 + C\alpha )^2} & \Big(
\big(G^{(\alpha)}_{m^1,n^1,n^1,m^1}-
G^{(\alpha)}_{0,n^1,n^1,0}\big) G^{(\alpha)}_{m^2,n^2,n^2,m^2}
\nonumber
\\*
&+ G^{(\alpha)}_{0,n^1,n^1,0} \big( G^{(\alpha)}_{m^2,n^2,n^2,m^2}
- G^{(\alpha)}_{0,n^2,n^2,0}\big)\;.
  \label{Q0-slice-i}
\end{align}
Taking (\ref{th1}) into account, it remains to obtain estimations for 
$G^{(\alpha)}_{m,n,n,m}- G^{(\alpha)}_{0,n,n,0}$. From
(\ref{eq:propinit-b}) we have, expressing $m+h$ by $n$,
\begin{align}
&\big|G^{(\alpha)}_{m,n,n,m}- G^{(\alpha)}_{0,n,n,0}\big|
\nonumber
\\
&=\left(\frac{\sqrt{1-\alpha}}{1+C \alpha} \right)^{n} 
\left|
\left(\frac{\sqrt{1-\alpha}}{1+C \alpha} \right)^{m} 
\sum_{u=0}^{\min(m,n)} \binom{m}{u} \binom{n}{u} 
\left(\frac{C \alpha (1+\Omega)}{\sqrt{1-\alpha}\,(1-\Omega)} 
\right)^{2u}
- 1 \right|\;.
\label{Taylor-0}
\end{align}
The $(u=0)$-part of the sum and the separate term $-1$ yield
\begin{align}
1- \left(\frac{\sqrt{1-\alpha}}{1+C \alpha} \right)^{m}
&= \left(1- \frac{\sqrt{1-\alpha}}{1+C \alpha} \right) 
\sum_{j=0}^{m-1} \left(\frac{\sqrt{1-\alpha}}{1+C \alpha} \right)^j
\nonumber
\\*
& \les m   \left(1- \frac{\sqrt{1-\alpha}}{1+C \alpha} \right)  
\nonumber
\\*
& \les m \alpha(C+1) \;. 
\label{Taylor-1}
\end{align}
Next, using 
$\binom{m}{v+1}=\frac{m-v}{v+1} \binom{m}{v}$ and our central estimate 
$\binom{m}{v} \les \frac{m^v}{v!}$, we have 
\begin{align}
&\sum_{u=1}^{m} \binom{m}{u} \binom{n}{u} 
\left(\frac{C \alpha (1+\Omega)}{\sqrt{1-\alpha}\,(1-\Omega)} 
\right)^{2u}
\nonumber
\\*
& \les \sum_{v=0}^{m-1} \frac{m-v}{v+1} \binom{m}{v} 
\left(\frac{C \alpha (1+\Omega)}{\sqrt{1-\alpha}\,(1-\Omega)} 
\right)^{v+1}
\sum_{u=0}^m 
\binom{n}{u} 
\left(\frac{C \alpha (1+\Omega)}{\sqrt{1-\alpha}\,(1-\Omega)} \right)^{u}
\nonumber
\\*
& \les m \left(\frac{C \alpha (1+\Omega)}{\sqrt{1-\alpha}\,(1-\Omega)} 
\right)
\exp\left(\frac{C \alpha (1+\Omega)(m+n)}{\sqrt{1-\alpha}\,(1-\Omega)} 
\right)
\nonumber
\\*
& \les m\alpha \frac{(1-\Omega)^2}{4\Omega \sqrt{1-M^{-1}}}
\exp\left(\frac{\alpha (1-\Omega^2)(m+n)}{4\Omega\sqrt{1-\alpha}}\right)\;.
\label{Taylor-2}
\end{align}
Inserting
(\ref{Taylor-1}) and (\ref{Taylor-2}) back into (\ref{Taylor-0}) 
%
we obtain for $\alpha \les M^{-1}$ the estimation
\begin{align}
\big|G^{(\alpha)}_{m,n,n,m}- G^{(\alpha)}_{0,n,n,0}\big|
&\les \alpha m \left(\frac{(1+\Omega)^2}{4\Omega} + 
\frac{(1-\Omega)^2}{4\Omega \sqrt{1-M^{-1}}} \right)
\nonumber
\\*
& \times 
\exp\left(\alpha n \left(
\frac{(1-\Omega^2)}{4\Omega\sqrt{1-\alpha}}
+ \frac{1}{\alpha} \ln \frac{4\Omega \sqrt{1-\alpha}}{
4\Omega+\alpha(1-\Omega)^2}\right)\right) \;.
\label{Delta-diff-0}
\end{align}
Comparing (\ref{Delta-diff-0}) with (\ref{th1-exp}) we obtain in 
(\ref{thcomp0}) the same restrictions to $\Omega$ as
in Theorem~\ref{thm-th1}.

To approach (\ref{thcomp1}) we consider
\begin{align}
\mathcal{Q}^{i(1)}_{\di{m^1}{m^2}, \di{n^1}{n^2}; 
\di{n^1}{n^2}, \di{m^1}{m^2}} 
&= \frac{\theta}{8\Omega} \int_{M^{-i}}^{M^{-i+1}} d\alpha\,  
\dfrac{(1-\alpha)^{\frac{\mu_0^2 \theta}{8 \Omega}}}{  
(1 + C\alpha )^2} 
\nonumber
\\*
& \times \Big(
\Big(G^{(\alpha)}_{m^1,n^1,n^1,m^1}-
G^{(\alpha)}_{0,n^1,n^1,0}
-m^1\big(G^{(\alpha)}_{1,n^1,n^1,1}-
G^{(\alpha)}_{0,n^1,n^1,0}\big)
\Big) G^{(\alpha)}_{m^2,n^2,n^2,m^2}
\nonumber
\\*
& \qquad + m^1\big(G^{(\alpha)}_{1,n^1,n^1,1}-
G^{(\alpha)}_{0,n^1,n^1,0}
\big) \big(G^{(\alpha)}_{m^2,n^2,n^2,m^2}
-G^{(\alpha)}_{0,n^2,n^2,0}\big)
\nonumber
\\*
&+ G^{(\alpha)}_{0,n^1,n^1,0} \Big( 
G^{(\alpha)}_{m^2,n^2,n^2,m^2}
- G^{(\alpha)}_{0,n^2,n^2,0}
-m^2\big(G^{(\alpha)}_{1,n^2,n^2,1}
-G^{(\alpha)}_{0,n^2,n^2,0}\big)
\Big)\Big)\;.
  \label{Q1-slice-i}
\end{align}
The third line is estimated by (\ref{Delta-diff-0}). In the
other lines we have for $n\ges 1$
\begin{align}
&\big|G^{(\alpha)}_{m,n,n,m}- G^{(\alpha)}_{0,n,n,0}
-m(G^{(\alpha)}_{1,n,n,1}- G^{(\alpha)}_{0,n,n,0})\big|
\nonumber
\\
&=\left(\frac{\sqrt{1-\alpha}}{1+C \alpha} \right)^{n} 
\Bigg|
\left(\frac{\sqrt{1-\alpha}}{1+C \alpha} \right)^{m} 
-m\left(\frac{\sqrt{1-\alpha}}{1+C \alpha} -1 \right)-1 \Bigg|
\nonumber
\\*
&+ mn \left(\frac{\sqrt{1-\alpha}}{1+C \alpha} \right)^{n+1} 
 \left(\frac{C \alpha (1+\Omega)}{\sqrt{1-\alpha}\,(1-\Omega)} 
\right)^{2}
\Bigg| \left(\frac{\sqrt{1-\alpha}}{1+C \alpha} \right)^{m-1} -1\Bigg|
\nonumber
\\*
&
+ \left(\frac{\sqrt{1-\alpha}}{1+C \alpha} \right)^{m+n} 
\sum_{u=2}^{\min(m,n)} \binom{m}{u} \binom{n}{u} 
\left(\frac{C \alpha (1+\Omega)}{\sqrt{1-\alpha}\,(1-\Omega)} 
\right)^{2u}\;.
\label{Taylor-3}
\end{align}
In the third line we use $n\frac{C \alpha
  (1+\Omega)}{\sqrt{1-\alpha}\,(1-\Omega)} \les \exp\left(
n\alpha \frac{(1-\Omega)^2}{4\Omega \sqrt{1-\alpha}}\right)$.
The further procedure is as before.

Finally, we consider
\begin{align}
&\mathcal{Q}^{i(\frac{1}{2})}_{\di{m^1+1}{m^2}, \di{n^1+1}{n^2}; 
\di{n^1}{n^2}, \di{m^1}{m^2}} 
\nonumber
\\*
&= \frac{\theta}{8\Omega} \int_{M^{-i}}^{M^{-i+1}} d\alpha\,  
\dfrac{(1-\alpha)^{\frac{\mu_0^2 \theta}{8 \Omega}}}{  
(1 + C\alpha )^2}  \Big(
\big(G^{(\alpha)}_{m^1+1,n^1+1,n^1,m^1}-\sqrt{m^1+1}
G^{(\alpha)}_{1,n^1+1,n^1,0}\big) G^{(\alpha)}_{m^2,n^2,n^2,m^2}
\nonumber
\\*
&\hspace*{10em} 
+ \sqrt{m^1+1}
G^{(\alpha)}_{1,n^1+1,n^1,0} \big( G^{(\alpha)}_{m^2,n^2,n^2,m^2}
- G^{(\alpha)}_{0,n^2,n^2,0}\big)\Big)\;.
  \label{Q2-slice-i}
\end{align}
The estimation for the second line of (\ref{Q2-slice-i}) is
immediately obtained from (\ref{Delta-diff-0}) and (\ref{th2}). In the
second line we have
 \begin{align}
&\big|G^{(\alpha)}_{m+1,n+1,n,m}- \sqrt{m+1} 
G^{(\alpha)}_{1,n+1,n,0}\big|
\nonumber
\\*
&=
\Bigg|
\left(\frac{\sqrt{1-\alpha}}{1+C \alpha} \right)^{m} 
\sum_{u=0}^{\min(m,n)} \frac{\sqrt{(m+1)(n+1)}}{u+1} 
\binom{m}{u} \binom{n}{u} 
\left(\frac{C \alpha (1+\Omega)}{\sqrt{1-\alpha}\,(1-\Omega)} 
\right)^{2u+1}
\nonumber
\\*
& \qquad\qquad\qquad\qquad - \sqrt{(m+1)(n+1)}\left(\frac{C \alpha
    (1+\Omega)}{\sqrt{1-\alpha}\,(1-\Omega)}  \right)
 \Bigg|\left(\frac{\sqrt{1-\alpha}}{1+C \alpha} \right)^{n+1} \;.
\end{align}
We use the estimation $\sqrt{n\frac{C \alpha
  (1+\Omega)}{\sqrt{1-\alpha}\,(1-\Omega)}} \les \exp\left(
\frac{1}{2} n\alpha \frac{(1-\Omega)^2}{4\Omega \sqrt{1-\alpha}}\right)$
and proceed along the same lines as before.

This finishes the proof. \hfill $\square$

\section{Perturbative power-counting}
\label{sec:powercounting}

\subsection{Ribbon graphs}

Consider a given $\phi^4$-ribbon graph $G$ with $N$ external legs, $V$
vertices, $I$ inner lines and $F$ faces, hence of genus $g= 1 - \frac 12
(V-I+F)$.  There are four indices $\{m,n;k,l\} \in \mathds{N}^2$
associated to each inner line $\parbox{11mm}{\begin{picture}(10,8)
    \put(0,3){\epsfig{file=p1,scale=1,bb=71 667 101 675}}
    \put(2,6.5){\mbox{\scriptsize$n$}}
    \put(6,1){\mbox{\scriptsize$l$}} \put(0,1){\mbox{\scriptsize$m$}}
    \put(8,6.5){\mbox{\scriptsize$k$}}
\end{picture}}$ of the graph and two indices for each external line, hence
$4 I +2N =8V$ such indices. But at every vertex, the left index of a
ribbon coincides with the right one of the next ribbon.  This creates
$4V$ independent identifications, so we can write the indices of any
propagator in terms of a set $\mathcal{I}$ of $4V$ indices, four per vertex,
for instance each ``left'' half-ribbon index.

The amplitude of the graph is then the sum
\begin{align}  
  A_{G} = \sum_{\mathcal{I}} \prod_{\delta \in G}
  G_{m_{\delta}(\mathcal{I}),n_{\delta}(\mathcal{I});k_{\delta}(\mathcal{I}),l_{\delta}(\mathcal{I})}\;
    \delta_{m_{\delta}-l_{\delta},n_{\delta}-k_{\delta}} \;,
\label{IG}
\end{align}
where the four basic indices of the propagator $G$ for a line $\delta$
are functions of $\mathcal{I}$ called $\{m_{\delta}(\mathcal{I}),n_{\delta}(\mathcal{I});
k_{\delta}(\mathcal{I}),l_{\delta}(\mathcal{I})\} $.

The scale decomposition of the propagator being
\begin{align}  
G = \sum_{i=0}^{\infty}G^i\;,
\end{align}
we have an associated decomposition of any amplitude of the theory as
\begin{align}  
A_G &= \sum_{\mu} A_{G,\mu}\;,
\\
A_{G,\mu} &= \sum_{\mathcal{I}} \prod_{\delta \in G} G^{i_{\delta}}_{
m_{\delta}(\mathcal{I}),n_{\delta}(\mathcal{I});
k_{\delta}(\mathcal{I}),l_{\delta}(\mathcal{I})}  \;
\delta_{m_{\delta}(\mathcal{I})-l_{\delta}(\mathcal{I}),
n_{\delta}(\mathcal{I})-k_{\delta}(\mathcal{I})} \;,
\label{IGmu}
\end{align}
where $\mu=\{i_{\delta}\}$ runs over all possible attributions of a
positive integer $i_{\delta}$ for each line $\delta$. Such a $\mu$ is
therefore called a scale attribution.

We recall our two main bounds on the propagator
\begin{align}
 G^{i}_{m,n;k,l} &\les K M^{-i} e^{-cM^{-i} (\|m\|+\|n\|+\|k\|+\|l\|)} \;,
\label{cos}
\\
\sum_{l}\max_{n,k}G^{i}_{m,n;k,l} &\les K' M^{-i}  e^{-c'M^{-i} \|m\|}\;,
\label{nocos}
\end{align}
for some constants $K,K'$ and $c,c'$.

A considerable fraction of the $4V$ indices initially associated to this graph
is determined by external indices of the graph and the $\delta$-function in
(\ref{IG}). The undetermined indices are summation indices. Perturbative power
counting for a graph consists in finding the indices for which the summation
costs a factor $M^{2i}$, and the ones for which it costs only
$\mathcal{O}(1)$, thanks to (\ref{nocos}). The factor $M^{2i}$ follows from
(\ref{cos}) after summation over some index\footnote{Remember that there are
  two symplectic pairs, one for spatial dimensions 1 and 2, and the other for
  spatial dimensions 3 and 4.} $m \in \mathds{N}^2$,
\begin{align}
\sum_{m^1,m^2=0}^\infty e^{- c M^{-i}(m^1+m^2)} = \frac{1}{(1-e^{- c
    M^{-i}})^2} = \frac{M^{2i}}{c^2} (1 + \mathcal{O}(M^{-i}))\;.
\end{align}

\subsection{Dual graphs}

We first want to use as much as possible the $\delta$-functions in
(\ref{IG}) to reduce the set $\mathcal{I}$ to a truly minimal set $\mathcal{I}'$ of
independent indices. It is convenient for this task to consider 
the dual graph for which the problem becomes analogous to an ordinary
problem of momentum routing.

The dual graph of a ribbon graph is obtained by associating to each
face a vertex and to each vertex a face. Every line bordering two
neighboring faces is replaced by a line joining the two corresponding
vertices of the dual graph. Hence, the genus does not change under
this duality.  We shall write $V'=F$, $F'=V$ for the number of
vertices and faces of the dual graph (dual quantities are
usually distinguished by a prime).  If the initial graph is a
$\phi^4$-graph, i.e.\ has coordination 4 at each vertex, we have
$4=I_{f'}+N_{f'}$ for each face $f' \in F'$, where $I_{f'}$ and
$N_{f'}$ denote the numbers of edges and external valences,
respectively, which belong to $f'$. The coordination at the vertices
of the dual graph is arbitrary.

The construction of the dual of a graph goes as follows: First, for
each oriented face of the original ribbon graph, draw an oriented
ribbon vertex by assigning
\begin{itemize}
\item to a single line of a propagator of the original graph an internal
valence of the dual vertex,
\item to an external valence of the original graph an external
valence of the dual vertex,
\end{itemize}
respecting the order according to the arrows on the trajectories. 
In the second step we connect the valences by the duals of the
propagators of the original graph, which is obtained according to 
\begin{align}
\parbox{4cm}{\epsfig{figure=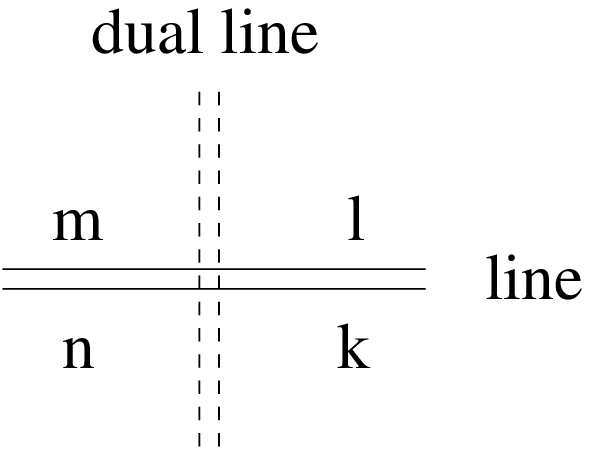,width=4cm,bb=0 0 170 128}}
\label{dualprop}
\end{align}

Let us consider the following example of a ribbon graph with a single face: 
\begin{align}
\parbox{42mm}{\begin{picture}(20,24)
       \put(0,0){\epsfig{file=a22a,bb=71 613 184 684}}
       \put(2,17){\mbox{\footnotesize$m_2$}}
       \put(0,11){\mbox{\footnotesize$n_2$}}
       \put(13,11){\mbox{\footnotesize$r'$}}
       \put(32,11){\mbox{\footnotesize$r$}}
       \put(32,22){\mbox{\footnotesize$m_1$}}
       \put(25,24){\mbox{\footnotesize$n_1$}}
       \put(13.5,17.5){\mbox{\footnotesize$q$}}
       \put(25,10.5){\mbox{\footnotesize$q'$}}
   \end{picture}}
\end{align}
The above rules lead to the following dual vertex: 
\begin{align}
\parbox{42mm}{\begin{picture}(40,40)
       \put(0,0){\epsfig{file=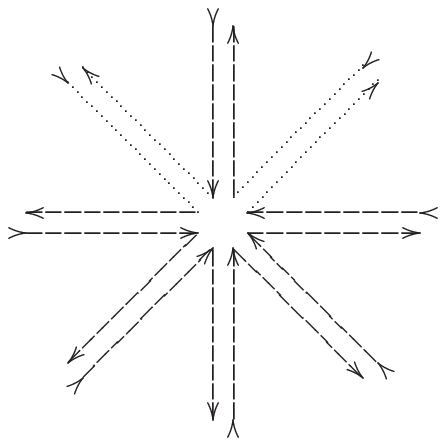,bb=71 569 186 684}}
       \put(3,23){\mbox{\footnotesize$n_1$}}
       \put(3,17){\mbox{\footnotesize$q$}}
       \put(6,10.5){\mbox{\footnotesize$q$}}
       \put(10,6){\mbox{\footnotesize$r$}}
       \put(16,4){\mbox{\footnotesize$r$}}
       \put(22,4){\mbox{\footnotesize$r'$}}
       \put(28,5){\mbox{\footnotesize$r'$}}
       \put(34,10){\mbox{\footnotesize$q'$}}
       \put(34,23){\mbox{\footnotesize$m_2$}}
       \put(35,16){\mbox{\footnotesize$q'$}}
       \put(32,29){\mbox{\footnotesize$m_2$}}
       \put(27,34){\mbox{\footnotesize$n_2$}}
       \put(14,37){\mbox{\footnotesize$m_1$}}
       \put(22,36){\mbox{\footnotesize$n_2$}}
       \put(4,30){\mbox{\footnotesize$n_1$}}
       \put(9,34){\mbox{\footnotesize$m_1$}}
     \end{picture}}
\end{align}     
Now we connect
the valences by the duals of the propagators of the original graph, i.e.\
$n_1q$ with $r'q'$, $qr$ with $q'm_2$ and $n_2m_1$ with $rr'$:
\begin{align}
\parbox{42mm}{\begin{picture}(40,44)
       \put(0,0){\epsfig{file=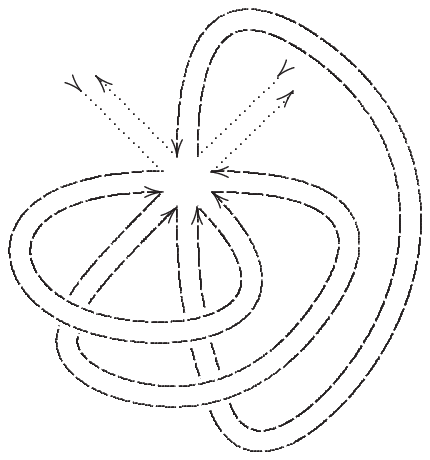,bb=76 548 197 677}}
       \put(9.5,31){\mbox{\footnotesize$n_1$}}
       \put(11,24.5){\mbox{\footnotesize$q$}}
       \put(15,20){\mbox{\footnotesize$r$}}
       \put(20.2,18.5){\mbox{\footnotesize$r'$}}
       \put(25,23){\mbox{\footnotesize$q'$}}
       \put(25,30){\mbox{\footnotesize$m_2$}}
       \put(21,36){\mbox{\footnotesize$n_2$}}
       \put(13,36){\mbox{\footnotesize$m_1$}}
\end{picture}}
\label{a22a-dual}
\end{align}   

The dual graph is made of the same propagators as the original graph,
only the index assignment is different. Whereas in the original graph
we have $G_{mn;kl}=\parbox{11mm}{\begin{picture}(10,8)
    \put(0,3){\epsfig{file=p1,scale=1,bb=71 667 101 675}}
    \put(2,6.5){\mbox{\scriptsize$n$}}
    \put(6,1){\mbox{\scriptsize$l$}} \put(0,1){\mbox{\scriptsize$m$}}
    \put(8,6.5){\mbox{\scriptsize$k$}}
  \end{picture}}$, the index assignment for
propagators in the dual graph is 
\begin{align}
G_{mn;kl} = \parbox{11mm}{\begin{picture}(10,8.5)
    \put(0,3){\epsfig{file=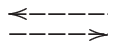,scale=1,bb=71 667 101 675}}
    \put(2,6.5){\mbox{\scriptsize$l$}}
    \put(6,1){\mbox{\scriptsize$n$}} \put(0,1){\mbox{\scriptsize$m$}}
    \put(8,6.5){\mbox{\scriptsize$k$}}
  \end{picture}}\;.
\label{dual-assign}
\end{align}
The conservation rule $\delta_{m-l,-(k-n)}$ in (\ref{IG}) now states
that the difference between outgoing and incoming indices of the
half-propagator attached to a dual vertex, namely $m-l$, is conserved
as minus the corresponding difference $k-n$ at the other end of the
propagator. Actually, these index differences describe the
\emph{angular momentum}, and the conservation of these differences
$\ell=m-l$ and $-\ell=k-n$ is nothing but the angular momentum
conservation due to the $SO(2)\times SO(2)$ symmetry of the
noncommutative $\phi^4$-action. Thus, taking the incoming indices as
the reference, the angular momentum through the dual propagator
determines the outgoing indices:
\begin{align}
\parbox{30mm}{\begin{picture}(30,14)
    \put(0,1){\epsfig{file=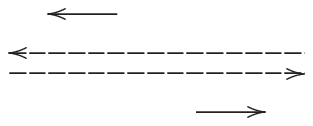,scale=1,bb=71 654 158 684}}
    \put(2,8.5){\mbox{\scriptsize$l$}}
    \put(26,3.5){\mbox{\scriptsize$n$}} 
    \put(1,3.5){\mbox{\scriptsize$m$}}
    \put(27,8.5){\mbox{\scriptsize$k$}}
    \put(8,12.5){\mbox{\scriptsize$\ell$}}
    \put(20,-1){\mbox{\scriptsize$-\ell$}}
  \end{picture}}\qquad\qquad 
m=l+\ell\;,~~ k=n+(-\ell)\;.
\label{angmom}
\end{align}
In the same way, there are external angular momenta $\wp$ which enter
the dual graph through the external valences. We shall also use the
incoming arrow as the reference so that the angular momentum is the
index difference between outgoing and incoming arrows. Furthermore, by
cyclicity at any vertex, the sum of all incoming differences at a
vertex, i.e.\ the sum of incoming angular momenta, must be zero. Of
course, this implies that the total external angular momentum entering
a graph is zero, too.

Thus, the angular momentum for the dual graph is exactly like the
usual momentum in an ordinary Feynman graph: a momentum that goes out
like $p$ at one half-end, must go out as $-p$ at the other half-end,
and the total momentum is conserved at any vertex. (In Feynman graphs
this follows from translation invariance.)  It should be stressed that
one has to take into account positivity constraints for the angular
momenta $\ell$: they lie in $\mathds{Z}$, but all indices $m,n,k,l$
must be {\it positive} integers.

We therefore know that the number of {\it independent} index 
differences is exactly the number of {\it loops} $L'$ of the dual
graph. For a connected graph, this number is $L'= I-V'+1$.
Furthermore, each index at a vertex is clearly only a function of the
differences at a vertex and of a single \emph{reference index} for the
dual vertex. If the dual vertex is an external one, we take as the
reference index the \emph{outgoing} index at one of the external legs.
If the dual vertex is an internal one, we have to make a choice
(determined later) for the reference index. These internal vertex
reference indices correspond to the loop variable of the original
graph.  Therefore, after using the conservation rules or
$\delta$-functions of each propagator, the number of independent
indices to be summed for every graph is simply $V'-B + L' = I +
(1-B)$. Here, $B \ges 1$ is the number of boundary components of the
original graph, which coincides with the number of external vertices
of the dual graph.

Expressing each index in the graph as a function of a set $\mathcal{I}'$ of
such independent indices is therefore identical to the problem called
momentum routing in a Feynman graph. It is well-known that the
solution is not canonical or unique. A good way to root the momenta is
to pick a spanning tree $\mathcal{T}_\mu$ of the dual graph, with $V'-1$
lines, and to use the complement set $\mathcal{L}_\mu$ as the set of
fundamental independent differences.  The subscript $\mu$ refers to the
choice of the tree which depends on the scale attribution $\mu$ in
(\ref{IGmu}).

\subsection{Choice of the tree}

A given scale attribution $\mu = \{i_\delta\}$ defines an order of
lines in the dual graph. We define 
\begin{align}
\delta_1 \les \delta_2 \les \dots
\les \delta_I \qquad \text{if} \qquad 
i_{\delta_1} \les i_{\delta_2} \les \dots \les
i_{\delta_I}\;.
\label{order}
\end{align}
In case of equality we make any choice. Let $\delta_1^{\mathcal{T}}$
be the smallest line with respect to this order which is not a
tadpole, and
$G^{i_{\delta^{\mathcal{T}}_1}}_{m_{\delta^{\mathcal{T}}_1};
  n_{\delta^{\mathcal{T}}_1}; k_{\delta^{\mathcal{T}}_1},
  l_{\delta^{\mathcal{T}}_1}}$ be the corresponding propagator.  The
line $\delta^{\mathcal{T}}_1$ will then connect two vertices $v^\pm_1$
and forms the first segment of the tree. We let $\mu_1:= \mu\setminus
(\delta_1 \cup \dots \cup \delta^T_1)$ and $\mathcal{T}_1 =
\delta^{\mathcal{T}}_1 \cup v^+_1 \cup v^-_1$.

In the remaining set $\mu_1$ of lines we identify the smallest
line $\delta^{\mathcal{T}}_2$ of $\mu_1$ which does not form a loop
when added to $\mathcal{T}_1$. We define $\mu_2= \mu \setminus
(\delta_1 \cup \dots \cup \delta^T_2)$ and
\begin{itemize}
\item $\mathcal{T}_2 = \mathcal{T}_1 \cup \delta^{\mathcal{T}}_2 \cup 
v^+_2$ if $\delta^{\mathcal{T}}_2$ connects a vertex $v^+_2$ to 
$\mathcal{T}_1$,

\item  $\mathcal{T}_2 = \mathcal{T}_1 \cup \delta^{\mathcal{T}}_2 \cup 
v^+_2 \cup v^-_2$ if $\delta^{\mathcal{T}}_2$ connects two vertices 
$v^\pm_2 \notin \mathcal{T}_1$.
\end{itemize}

In the $n^{\text{th}}$ step, we identify the smallest
line $\delta^{\mathcal{T}}_n$ of $\mu_{n-1}$ which does not form a loop
when added to $\mathcal{T}_{n-1}$. We define $\mu_n= \mu \setminus
(\delta_1 \cup \dots \cup \delta^T_n)$ and
\begin{itemize}
\item $\mathcal{T}_n = \mathcal{T}_{n-1} \cup \delta^{\mathcal{T}}_n \cup 
v^+_n$ if $\delta^{\mathcal{T}}_2$ connects a vertex $v^+_n$ to 
$\mathcal{T}_{n-1}$,

\item  $\mathcal{T}_n = \mathcal{T}_{n-1} \cup \delta^{\mathcal{T}}_n \cup 
v^+_n \cup v^-_n$ if $\delta^{\mathcal{T}}_2$ connects two vertices 
$v^\pm_n \notin \mathcal{T}_{n-1}$,

\item  $\mathcal{T}_n = \mathcal{T}_{n-1} \cup \delta^{\mathcal{T}}_n$ 
if $\delta^{\mathcal{T}}_2$ connects two disjoint subsets of 
$\mathcal{T}_{n-1}$.
\end{itemize}
The $(V'-1)^{\text{th}}$ step leads to the desired tree $\mathcal{T}_\mu =
\mathcal{T}_{V'-1}$. The importance of this construction is the fact that
any line $\delta^{\mathcal{L}}_j \in \mathcal{L}_\mu$ which connects
\emph{different} vertices $v^{\pm}_j$ of the tree has a scale index
$i_{\delta_j^{\mathcal{L}}}$ which is \emph{not smaller} than any
scale index of the lines \emph{in the tree} which connect $v^{\pm}_j$.
Such a tree optimization for a given scale attribution is one of
the most basic tools of constructive field 
theory \cite{Rivasseau:1991ub}, so it is
an encouraging sign that it appears also here in a natural way.

In the graphical representations we distinguish the tree by triple
dashed lines.

\subsection{Index assignment for a tree}

For the previously constructed tree $T_\mu$ we select one of its
$B \ges 1$ external vertices as the root $v_0$ of the tree. We relabel
the vertices in the tree such that all vertices in the subtree above a
vertex $v_n$ have a label bigger than $n$. 

The order (\ref{order}) of the lines of the graph provides us with a
convenient position for the main reference index $m$ at each vertex.
If $v$ is an internal vertex, we let $\delta_v$ be the smallest line
in (\ref{order}) which is attached to $v$. By construction of the tree
we know that either $\delta_v$ is a tadpole, or it belongs to the
tree. We choose the \emph{outgoing} index (without the arrow when
viewed from the vertex) of this particular line $\delta_v$ as the main
reference index $m_v$. We let $\mathcal{G}_{\mathcal{M}}$ be the set of
lines at which a main reference index resides. It is possible that both
outgoing indices of a line $\delta_v=\delta_{v'}$ attached to $v$ and
$v'$ are main reference indices. In this case we let
$\delta_v$ appear twice in $\mathcal{G}_{\mathcal{M}}$. Thus,
$\mathcal{G}_{\mathcal{M}}$ consists of $V'-B$ elements.
If $v$ is an external vertex, we take as the
``main reference index'' the outgoing index at any external leg. The
following graph shows a typical situation of a tree and its main
reference indices, assuming absence of tadpoles and $B=1$:
\begin{align}
\parbox{60mm}{\begin{picture}(60,37)
\put(0,0){\epsfig{file=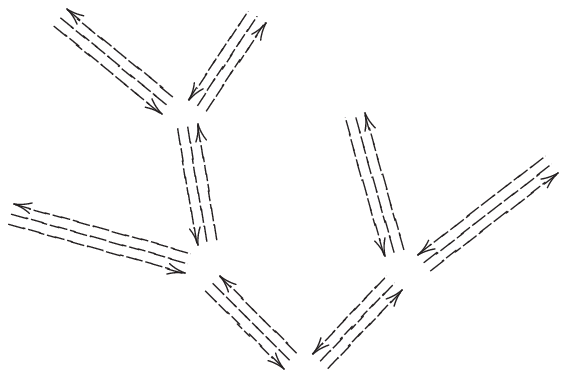,bb=71 578 233 684}}
\put(29,-1){\mbox{\scriptsize$v_0$}}
\put(19.5,11){\mbox{\scriptsize$v_1$}}
\put(22.5,14){\mbox{\scriptsize$m_{v_1}$}}
\put(39,10){\mbox{\scriptsize$v_2$}}
\put(40,16){\mbox{\scriptsize$m_{v_2}$}}
\put(-3,16){\mbox{\scriptsize$v_3$}}
\put(0,12){\mbox{\scriptsize$m_{v_3}$}}
\put(16.5,26){\mbox{\scriptsize$v_4$}}
\put(11.5,22.5){\mbox{\scriptsize$m_{v_4}$}}
\put(2.5,36.5){\mbox{\scriptsize$v_5$}}
\put(3,32){\mbox{\scriptsize$m_{v_5}$}}
\put(25,37){\mbox{\scriptsize$v_6$}}
\put(18,36){\mbox{\scriptsize$m_{v_6}$}}
\put(34,27){\mbox{\scriptsize$v_7$}}
\put(29,23){\mbox{\scriptsize$m_{v_7}$}}
\put(56,21.5){\mbox{\scriptsize$v_8$}}
\put(48,22){\mbox{\scriptsize$m_{v_8}$}}
\end{picture}}
\label{mainindices}
\end{align}

One can now write down every index in a unique way in terms of 
\begin{itemize}
\item $V'-B$ main reference indices $m_v$,

\item $L'$ internal angular momenta $\ell_\delta$ for
the set $\mathcal{L}_\mu$ of ``loop lines'' 

\item $B$ main reference indices at external vertices,

\item $N$ external angular momenta $\wp_\epsilon$ for the set
  $\mathcal{N}$ of external lines. 
\end{itemize}
The rule goes as follows.  One writes first the indices for all the
``leaves'' of the tree, that is the vertices (distinct from the root)
with coordination 1 in the tree (i.e.\ vertices $v_3,v_5,v_6,v_7$ and
$v_8$ in (\ref{mainindices})). For them, starting from the main index
$m_v$ (at the left of the unique line of $\mathcal{T}_\mu$ at $v$ that
goes towards the root, unless a tadpole at $v$ has the smallest
scale), which agrees with the incoming index of the next line at $v$
in clockwise direction, we compute all other indices by turning
\emph{clockwise} around the vertex and by adding the angular momenta
(internal or external ones) associated according to (\ref{angmom})
with the loop lines $\delta_1,\dots,\delta_k$ and possibly external
lines $\epsilon_1,\dots,\epsilon_{k'}$.  This gives indices $m_v +
\ell_1$, $m_v + \ell_1 + \ell_2$, \dots until we arrive at $m_v +
\ell_1 + \dots + \ell_{k+k'}$ which is at the right of the unique
line at $v$ that goes towards the root. Some of the $\ell_j$ could be
external angular momenta. Then we can compute the angular momentum
associated to that line: it is simply $-(\ell_1 + \dots +
\ell_{k+k'})$.  The following picture shows these assignments for a
particular ``leaf'':
\begin{align}
\parbox{60mm}{\begin{picture}(60,39)
\put(0,0){\epsfig{file=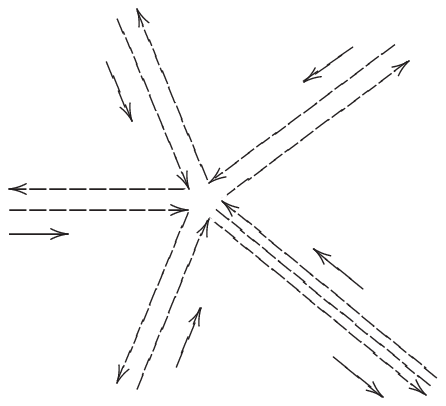,bb=71 570 198 684}}
\put(20,14){\mbox{\scriptsize$m_v$}}
\put(19,5){\mbox{\scriptsize$\ell_1$}}
\put(7,16.5){\mbox{\scriptsize$m_v{+}\ell_1$}}
\put(2,14){\mbox{\scriptsize$\ell_2$}}
\put(2,23){\mbox{\scriptsize$m_v{+}\ell_1{+}\ell_2$}}
\put(8,30){\mbox{\scriptsize$\ell_3$}}
\put(17,33){\mbox{\scriptsize$m_v{+}\ell_1$}}
\put(19,30){\mbox{\scriptsize${+}\ell_2{+}\ell_3$}}
\put(31,36){\mbox{\scriptsize$\ell_4$}}
\put(25,20){\mbox{\scriptsize$m_V{+}\ell_1{+}\ell_2{+}\ell_3{+}\ell_4$}}
\put(34,15){\mbox{\scriptsize$-(\ell_1{+}\ell_2{+}\ell_3{+}\ell_4)$}}
\put(18,0){\mbox{\scriptsize$\ell_1{+}\ell_2{+}\ell_3{+}\ell_4$}}
\end{picture}}
\label{leaf}
\end{align}
Having done this for all leaves, we can prune these leaves and
consider the next layer of vertices down towards the root (i.e.\
vertices $v_2$ and $v_4$ in (\ref{mainindices})) and reiterate
the argument.

Any summation index at a vertex $v$ is now clearly equal to $m_v$ plus a
linear combination of angular momenta $\ell_\delta$ for the set of lines
$\mathcal{L}_v \cup \mathcal{N}_v$ hooked to the ``subtree above $v$'', that
is the lines hooked at least at one end to a vertex $v'$ such that the unique
path from $v'$ to $v_0$ passes through $v$.

We split therefore the set $\mathcal{I}'$ of independent indices to be
summed into the two sets: 
\begin{itemize}
\item the set $\mathcal{M}_\mu =\{m_v\}$ of main reference indices at
  inner vertices, which consists of $V'-B$ elements,

\item the set
$\mathcal{J}_\mu= \{\ell_\delta,\; \delta \in \mathcal{L}\}$ of
angular momenta, which consists of $L'$ elements.  
\end{itemize}
The amplitude of
the graph $G$ is now written as:
\begin{align} 
  A_G &= \sum_{\mu} \sum_{\mathcal{M}_\mu, \mathcal{J}_\mu} 
\prod_{\delta \in G}
  G^{i_{\delta}}_{m_{\delta}(\mathcal{M}_\mu,\mathcal{J}_\mu),
n_{\delta}(\mathcal{M}_\mu,\mathcal{J}_\mu);
k_{\delta}(\mathcal{M}_\mu,\mathcal{J}_\mu),
l_{\delta}(\mathcal{M}_\mu,\mathcal{J}_\mu)}\; 
\chi(\mathcal{M}_\mu,\mathcal{J}_\mu)\;,
\label{AG}
\end{align}
where the sum over $\mathcal{M}_\mu$ is over positive integers, but
the sum over $\mathcal{J}_\mu$ is over relative integers and the
function $\chi(\mathcal{M}_\mu,\mathcal{J}_\mu)$ is the characteristic
function which states that all the functions
$\{m_{\delta}(\mathcal{M}_\mu,\mathcal{J}_\mu),
n_{\delta}(\mathcal{M}_\mu,\mathcal{J}_\mu);
k_{\delta}(\mathcal{M}_\mu,\mathcal{J}_\mu),
l_{\delta}(\mathcal{M}_\mu,\mathcal{J}_\mu)\}$ are positive. The
dependence of $A_G$ on the external indices ($B$ reference indices at
external vertices and $N$ external angular momenta) is not made
explicit.

\subsection{Power-counting}

Our goal is now to prove that sums over difference indices can always
be performed through (\ref{nocos}), hence at no cost, {\it using
  precisely the propagator $G^{i_\delta}$ to perform the sum over
  the angular momentum $\ell_\delta$}. However, as the angular momenta
are entangled, we need appropriate maximizations of the other
$G^{i_\delta'}$ over $\ell_\delta$. These maximizations require
a carefully chosen order. For processing the angular momenta, all
main reference indices $\mathcal{M}_\mu$ and the external indices are
kept constant.

We start with the highest labelled leaf $v=v_{V'-1}$ according to the
previously defined order of vertices. Let this vertex carry loop lines
with corresponding index assignments according to (\ref{leaf}). We
first assume that $\delta_4$ is not a tadpole. Then, we should sum
over $\ell_1,\ell_2,\ell_3$ \emph{after} the $\ell_4$-summation, i.e.\
$\ell_1,\ell_2,\ell_3$ are constant with respect to the
$\ell_4$-summation. This guarantees that at least one side of the
lines $\delta_1,\delta_2,\delta_3$, namely the side attached to $v$,
is independent of $\ell_4$. We would thus maximise the lines in the
tree over $\ell_4$, but possibly also the $\ell_4$-dependence of the
other ends of $\delta_1,\delta_2,\delta_3$. Looking e.g.\ at
$\delta_3$ which connects $v$ with $v'$, the corresponding propagator
would be
\begin{align}
G^{i_3}_{m_v+\ell_1+\ell_2, k_{v'}-\ell_3;
  k_{v'},m_v+\ell_1+\ell_2+\ell_3}\;,
\end{align}
see (\ref{dual-assign}). Note that the incoming index viewed from $v'$
is always the reference index, labelled $k_{v'}$ in this case, and the
corresponding outgoing index is obtained by adding the opposite
angular momentum $-\ell_3$. The reference index $k_{v'}$ is either
independent of $\ell_4$ or $k_{v'}=k^0_{v'} \pm \ell_4$ for $k^0_{v'}$
being independent of $\ell_4$. Thus, the maximization of
$G^{i_3}$ over $\ell_4$ and possibly over other angular momenta
$\ell_j$, $j>4$, is precisely of the structure in
(\ref{nocos}). Later, it will be essential that concerning the
$\ell_3$-summation to be applied to $G^{i_3}$ we keep $\ell_2$
and $\ell_1$ constant, because $k_{v'}$ might also depend on $\ell_1$
and/or $\ell_2$. After having maximized all other propagators over
$\ell_4$ we restrict the $\ell_4$-summation to the line $\delta_4$,
which connects $v$ to $v''$ and corresponds to the propagator
\begin{align}
G^{i_4}_{m_v+\ell_1+\ell_2+\ell_3, k_{v''}-\ell_4;
  k_{v''},m_v+\ell_1+\ell_2+\ell_3+\ell_4}\;.
\label{Deltai4}
\end{align}
Here, there might be previous maximizations of $k_{v''}$ over
$\ell_j$, $j>4$, which we can bound by the maximization of
(\ref{Deltai4}) over all $k_{v''}$. To that maximized propagator we
apply for given $\ell_1,\ell_2,\ell_3$ the $\ell_4$-summation. Later,
when applying in the same manner the $\ell_j$-summations, $j=1,2,3$, to 
$G^{i_j}$, we have to maximise the previously processed
$G^{i_4}$ over $\ell_1,\ell_2,\ell_3$. In other words, the line
$\delta_4$ and the $\ell_4$-summation taken at the correct place
yield,  see (\ref{nocos}),
\begin{align}
\max_{\ell_1,\ell_2,\ell_3} \left( \sum_{\ell_4} \left(
    \max_{\ell_j,\,j>4} 
G^{i_4}_{m_v+\ell_1+\ell_2+\ell_3, k_{v''}(\ell_j)-\ell_4;
  k_{v''}(\ell_j),m_v+\ell_1+\ell_2+\ell_3+\ell_4}\right)\right) \les
K M^{-i_4}\;.
\label{Deltai4max}
\end{align}

To summarize, we can estimate the $\ell_j$-summations restricted to
the line $\delta_j$ \emph{at the highest-labelled leaf} by
(\ref{nocos}) if all angular momenta $\ell_k$ associated with those
lines $\delta_k$ which are encountered \emph{before} $\delta_j$ when
turning \emph{clockwise} around the leaf (starting from the main
reference index) are kept constant. It is not difficult to convince
oneself that the same rule also holds for tadpoles.

Now we remove the highest-labelled vertex $v_{V'-1}$ and look at
the highest labelled leaf $v_{V'-2}$ of the reduced tree
$\mathcal{T}\setminus v_{V'-1}$. Let it be again represented by
(\ref{leaf}). We continue to label the lines at $v_{V'-2}$ in
clockwise order from the distinguished main reference index. Lines
which belong to $\mathcal{T}$ and lines attached to $v_{V'-1}$
are left out, because their angular momenta are already identified.
These previous angular momenta can be considered as \emph{fixed}
external ones which are summed \emph{after} the new angular momenta at
$v_{V'-2}$.  Clearly, the same rule as above, namely to sum
\emph{later} over those angular momenta which are encountered
\emph{earlier} in clockwise order, allows us to use (\ref{nocos}) for
the summation over the new angular momenta at $v_{V'-2}$.

We repeat this procedure until we arrive at the root $v_0$. The result
is a bound for the $\mathcal{J}_\mu$-summation in (\ref{AG}).  There,
all propagators which correspond to tree-lines $\delta \in
\mathcal{T}_\mu$ are maximized over $\mathcal{J}_\mu$. For tree lines 
where all indices depend on $\mathcal{J}_\mu$ 
the bound, due (\ref{cos}), is given by
\begin{align}
\max_{\mathcal{J}_\mu} G^{i_\delta}_{m_{\delta}(\mathcal{M}_\mu,
\mathcal{J}_\mu),n_{\delta}(\mathcal{M}_\mu,\mathcal{J}_\mu);
k_{\delta}(\mathcal{M}_\mu,\mathcal{J}_\mu),
l_{\delta}(\mathcal{M}_\mu,\mathcal{J}_\mu)} 
\les K M^{-i_\delta} \;, 
\qquad \delta \in \mathcal{T}_\mu\;,~~\delta \notin \mathcal{G}_\mu\;.
\label{DT-1}
\end{align}
If one of the indices of $\delta \in \mathcal{T}_\mu$ is a main
reference index at $v$, we have 
\begin{align}
\max_{\mathcal{J}_\mu} G^{i_\delta}_{m_v,
n_{\delta}(\mathcal{M}_\mu,\mathcal{J}_\mu);
k_{\delta}(\mathcal{M}_\mu,\mathcal{J}_\mu),
l_{\delta}(\mathcal{M}_\mu,\mathcal{J}_\mu)} 
\les K M^{-i_\delta} e^{-c M^{-i_\delta} \|m_{v}\|}\;, 
\qquad \delta \in \mathcal{T}_\mu\;,~~\delta \in \mathcal{G}_\mu\;.
\label{DT-2}
\end{align}
If two indices of $\delta$ are main reference indices
at $v,v'$, we have
\begin{align}
\max_{\mathcal{J}_\mu} G^{i_\delta}_{m_v,
n_{\delta}(\mathcal{M}_\mu,\mathcal{J}_\mu);
m_{v'},l_{\delta}(\mathcal{M}_\mu,\mathcal{J}_\mu)} 
\les K M^{-i_\delta} e^{-c M^{-i_\delta} (\|m_{v}\|+\|m_v{'}\|)}\;, 
\qquad \delta \in \mathcal{T}_\mu\;,~~\delta \in \mathcal{G}_\mu
\setminus \delta\;.
\label{DT-3}
\end{align}
Next, each summed propagator which corresponds to a line
$\delta \in \mathcal{L}_\mu$ delivers according to (\ref{Deltai4max})
a factor $K M^{-i_\delta}$,
\begin{align}
\max_{\ell_1,\dots,\ell_{\delta-1}} \left( \sum_{\ell_\delta} \left(
    \max_{\ell_{\delta+1},\dots,\ell_{L'}} 
G^{i_\delta}_{m_\delta(\mathcal{J}_\mu\setminus \ell_\delta), 
n_\delta(\mathcal{J}_\mu\setminus \ell_\delta)-\ell_\delta;
k_\delta(\mathcal{J}_\mu\setminus \ell_\delta),
l_\delta(\mathcal{J}_\mu\setminus \ell_\delta)+\ell_\delta}
\right)\right) 
&\les K' M^{-i_\delta} \;,
\nonumber
\\*
\delta \in \mathcal{L}_\mu\;,~~ & \delta \notin \mathcal{G}_\mu\;.
\label{DL-1}
\end{align}
In the case that $\delta_j \in \mathcal{L}_\mu$
is a tadpole at $v$ which has the smallest scale index among the set
of lines at $v$ we obtain from (\ref{nocos}) the bound 
\begin{align}
\max_{\ell_1,\dots,\ell_{\delta-1}} \left( \sum_{\ell_\delta} \left(
    \max_{\ell_{\delta+1},\dots,\ell_{L'}} 
G^{i_\delta}_{m_v, 
n_\delta(\mathcal{J}_\mu\setminus \ell_\delta)-\ell_\delta;
k_\delta(\mathcal{J}_\mu\setminus \ell_\delta),
l_\delta(\mathcal{J}_\mu\setminus \ell_\delta)+\ell_\delta}
\right)\right) &\les
K' M^{-i_\delta} \,e^{-c' M^{-i_\delta} \|m_{v}\|}\;,
\nonumber
\\*
\delta \in \mathcal{L}_\mu\;,~~ &\delta \in \mathcal{G}_\mu\;.
\label{DL-2}
\end{align}
Eventually, there will be indices $m_\epsilon,n_\epsilon$ which are
fixed as external ones. Each one delivers according to (\ref{cos}) an
additional factor $e^{-cM^{-i_\epsilon}\|m_\epsilon\|}$ and
$e^{-cM^{-i_\epsilon}\|n_\epsilon\|}$, respectively, because these
decays cannot be removed by maximizing loop momenta. For external
indices which are not connected to internal lines we put
$c\equiv 0$.

Altogether, the $\mathcal{J}_\mu$-summation in (\ref{AG}) can be
estimated by
\begin{align}
 A_G &\les \sum_{\mu} \sum_{m_1,\dots,m_{V'-B} \in \mathds{N}^2} 
\left(\prod_{\delta' \in \mathcal{G}_\mu}
e^{-c M^{-i_{\delta'}} \|m_{v(\delta')}\|}\right)
\left(\prod_{\delta \in G}  K M^{-i_{\delta}} \right) 
\nonumber
\\*
& \qquad \times 
\left(\prod_{\epsilon=1}^N e^{-cM^{-i_\epsilon}\|m_\epsilon\|}\right)
\left(\prod_{\epsilon=1}^N e^{-cM^{-i_\epsilon}\|n_\epsilon\|}\right)\;,
\label{AG-1}
\end{align}
where $m_{v(\delta')}$ is the main reference index at $\delta' \in
\mathcal{G}_\mu$. After summation over $m_1,\dots,m_{V'-B}$
we have
\begin{align}
 A_G &\les \sum_{\mu} \frac{K^I}{c^{2(V'-B)}} 
\Big(M^{-\sum_{\delta \in G} i_{\delta}} \Big)
\Big(M^{2 \sum_{\delta' \in \mathcal{G}_\mu} i_{\delta'}}\Big)
\left(\prod_{\epsilon=1}^N e^{-cM^{-i_\epsilon}\|m_\epsilon\|}\right)\!
\left(\prod_{\epsilon=1}^N e^{-cM^{-i_\epsilon}\|n_\epsilon\|}\right).
\label{AG-2}
\end{align}

The dangerous region of the sum over the scale attribution is at large
scale indices. To identify this region, we associate to the order
(\ref{order}) of lines a sequence of subgraphs $G_I \subset G_{I-1}
\subset \dots \subset G_1=G$ of the original ribbon graph by defining
$G_\gamma$ as the possibly disconnected set of lines $\delta_{\gamma'}
\ges \delta_\gamma$, together with all vertices attached to them. To
$G_\gamma$ we associate the scale attribution $\mu_\gamma$ which
starts from an irrelevant low-scale cut-off $i_{\gamma-1}\les
i_{\gamma'}$. We conclude from (\ref{AG-2}) that the amplitude
$A_{G_\gamma}$ corresponding to the subgraph $G_\gamma$ diverges if 
\begin{align}
\omega_\gamma := 
2(V'_\gamma-B_\gamma)-I_\gamma=2F_\gamma-2B_\gamma-I_\gamma 
=(2-\tfrac{N_\gamma}{2})-2(2g_\gamma+B_\gamma-1)
\end{align}
is non-negative, where $N_\gamma$, $V_\gamma$, $I_\gamma=I-\gamma+1$,
$F_\gamma$ and $B_\gamma$ are the numbers of external legs, vertices,
edges, faces and external faces of $G_\gamma$, respectively, and
$g_\gamma= 1 - \frac{1}{2} (V_\gamma-I_\gamma+F_\gamma)$ is its
genus. We have thus proven the following
\begin{theorem}
\label{pc-slice}
  The sum over the scale attribution $\mu$ in (\ref{AG-2}) converges
  if for all subgraphs $G_\gamma \subset G$ we have $\omega_\gamma <0$.
\end{theorem}
For the total graph $\gamma=G$ the power-counting degree becomes
$\omega=(2-\frac{N}{2})-2(2g+B-1)$, which reproduces the
power-counting degree derived in \cite{Grosse:2003aj}.

We consider in Appendix~\ref{app:examples} a few examples for the sum over the
scale attribution.

\subsection{Subtraction procedure for divergent subgraphs}

The power-counting theorem \ref{pc-slice} implies that planar subgraphs with
two or four external legs are the only ones for which the sum over the scale
attribution can be divergent. These graphs require a separate analysis. We
first see from Theorem~\ref{thm-th2} that 
\begin{itemize}
\item only those planar four-leg subgraphs with \emph{constant index} along 
the trajectory are marginal,

\item only those planar two-leg graphs with \emph{constant index} along the
  trajectory are relevant,

\item only those planar two-leg graphs with an \emph{accumulated index jump
    of 2} along the trajectory are marginal.
\end{itemize}
For the other types of graphs there is a sufficient power of $M^{-i}$ through
the terms $(M^{-i} l)^\delta$ in (\ref{lowindex}) which makes the sum over the
scale attribution convergent.

For the remaining truly divergent graphs one performs similarly as in the 
BPHZ scheme a Taylor subtraction about vanishing external indices. For
instance, a marginal four-leg graph with amplitude $A_{mn;nk;kl;lm}$ is
written as 
\begin{align}
  A_{mn;nk;kl;lm}=  (A_{mn;nk;kl;lm}-  A_{00;00;00;00})+ A_{00;00;00;00}\;.
\end{align}
The difference of graphs $A_{mn;nk;kl;lm}- A_{00;00;00;00}$ can be expressed
as a linear combination involving the \emph{composite propagators}
(\ref{comp-0}). See also \cite{Grosse:2004yu} for more details. Then, the
estimation (\ref{thcomp0}) provides an additional factor $M^{-i}$ which makes
the sum over the scale attribution for
the difference $A_{mn;nk;kl;lm}- A_{00;00;00;00}$ convergent. 

Eventually, there remain only the four divergent base functions 
$A_{\di{0}{0}\di{0}{0};\di{0}{0}\di{0}{0};\di{0}{0}\di{0}{0};
\di{0}{0}\di{0}{0}}$, $A_{\di{0}{0}\di{0}{0};\di{0}{0}\di{0}{0}}$, 
$(A_{\di{1}{0}\di{0}{0};\di{0}{0}\di{1}{0}}
-A_{\di{0}{0}\di{0}{0};\di{0}{0}\di{0}{0}})$ and 
$A_{\di{1}{0}\di{1}{0};\di{0}{0}\di{0}{0}}$, taking into account the symmetry
properties of the model. These are normalized to their ``experimentally''
determined values: the physical coupling constant, the physical mass, 
the physical field amplitude and the physical frequency of the 
harmonic oscillator potential, respectively. 

At the end, any graph appearing in the noncommutative $\phi^4$-model has an
amplitude which is uniquely expressed by four normalization conditions as well
as convergent sums over the scale attribution. Thus, the model is
renormalizable to all orders.

\section{Conclusion}

For many years, noncommutative quantum field theories were supposed to be
ill-behaved due to the UV/IR-mixing problem \cite{Minwalla:1999px}.
Meanwhile, it turned out \cite{Grosse:2003aj,Grosse:2004yu} that at least the
Euclidean noncommutative $\phi^4_4$-model is as good as its commutative
version: it is renormalizable to all orders. In fact, the noncommutative
$\phi^4_4$-model is even better than the commutative version with respect to
one important issue: the behavior of the $\beta$-function. 

It is well-known that the main obstacle to a rigorous construction of the
commutative $\phi^4_4$-model is the non-asymptotic freedom of the theory. The
noncommutative model is very different: The computation of the
$\beta$-function \cite{Grosse:2004by} shows that the ratio of the bare
coupling constant to the square of the bare frequency parameter remains
(at the one loop level) constant over all scales,
$\frac{\lambda}{\Omega^2}=\text{const}$. (This was noticed in
\cite{Grosse:2004ik}.) As the bare frequency is bounded by $1$, this means
that the bare coupling constant is \emph{bounded}. For appropriate 
renormalized values, the coupling constant can be kept arbitrarily
small throughout the renormalization flow. We are, therefore, optimistic that a
rigorous construction of the noncommutative $\phi^4_4$-model will be
possible. 

In this paper we have undertaken the first important steps in this
direction.  We have formulated the perturbative renormalization proof
in a language which admits a direct extension to constructive methods.
More details about our program are given in \cite{Rivasseau:2004az}.
Moreover, our new renormalization proof is much more efficient than
the previous one (by a factor of 3 when looking at the number of
pages). Eventually, we have established analytical bounds for the
asymptotic behavior of the propagator which before were only
established numerically.

\begin{appendix}

 \section{The case $\Omega = 1$} \label{app1}

Our proofs do not apply to the case $\Omega=1$, because the scale parameter
becomes $M=1$, see (\ref{M-1}) and (\ref{Rbeta}). However, the case $\Omega=1$
can be directly treated. According to (\ref{eq:propinit-b}), only the 
terms with $u=m=l$ survive:  
\begin{align}
G^{\Omega=1}_{mn;kl}&=\frac{\theta}{8} 
\int_0^1d\alpha (1-\alpha)^{\frac{\mu_0^2\theta}{8} + (\frac{D}{4}-1)+ 
\frac{1}{2} (\|m\|+\|k\|)}\;\delta_{ml}\delta_{nk} 
\label{propfin}
\\*
&= \frac{\delta_{ml}\delta_{nk}}{
\mu_0^2+\frac{2}{\theta}(\|m\|+\|n\|+\|k\|+\|l\|+\frac{D}{2})}\;.
\end{align}
The exponential decay of the propagator in any index is
easily obtained from (\ref{propfin}) for all slices. Moreover, the $l$-sum 
is trivial to perform due to the index conservation $\delta_{ml}$ at each
trajectory.

\section{Examples for the sum over the scale attribution}
\label{app:examples}

We consider a few examples which underline the relation between
(\ref{AG-2}) and divergent subgraphs. The first one is:
\begin{align}
\parbox{50mm}{\begin{picture}(40,40)
\put(0,0){\epsfig{file=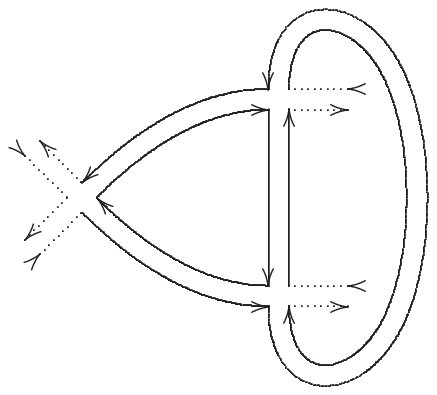,bb=71 571 189 682}}
\put(12,29){\mbox{\footnotesize$1$}}
\put(12,8){\mbox{\footnotesize$2$}}
\put(22,20){\mbox{\footnotesize$3$}}
\put(42,20){\mbox{\footnotesize$4$}}
\end{picture}}
\quad \mapsto \quad 
\parbox{40mm}{\begin{picture}(40,30)
\put(0,0){\epsfig{file=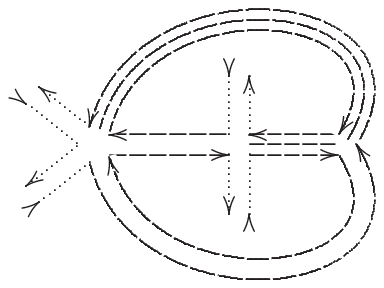,bb=71 591 173 672}}
\put(12,26){\mbox{\footnotesize$1$}}
\put(12,1){\mbox{\footnotesize$2$}}
\put(27,17){\mbox{\footnotesize$3$}}
\put(14,17){\mbox{\footnotesize$4$}}
\end{picture}}
\end{align}
The tree in the dual graph corresponds to $i_1\les i_2$ and $i_3\les 
i_4$ and $i_1\les i_4$. For the particular choice $i_1\les i_2\les i_3\le
i_4$ of the scale attribution we have to compute according to
(\ref{AG-2}) the following sum, taking into account that the volume
factor from the summation over the main reference index is associated
to $i_1$. We put $x=M^{-1}$, $0<x<1$:
\begin{align}
\sum_{i_4=0}^{\infty}\sum_{i_3=0}^{i_4}\sum_{i_2=0}^{i_3}
\sum_{i_1=0}^{i_2} x^{-i_1+i_2+i_3+i_4}
&= \sum_{i_4=0}^{\infty}\sum_{i_3=0}^{i_4}\sum_{i_2=0}^{i_3} 
\frac{1-x^{i_2+1}}{1-x} x^{i_3+i_4}
\nonumber
\\*
&= \sum_{i_4=0}^{\infty}\sum_{i_3=0}^{i_4}\Big( 
\frac{x^{i_3}-2 x^{i_3+1}}{(1-x)^2} 
+ i_3 \frac{x^{i_3}}{1-x} +
\frac{x^{2i_3+2}}{(1-x)^2} \Big) x^{i_4}
\nonumber
\\*
&= \sum_{i_4=0}^{\infty}\Big( \frac{
x^{i_4}-2x^{2i_4+1}+2 x^{2i_4+3}- x^{3i_4+4 }}{(1-x)^3(1+x)} 
- \frac{i_4\,x^{2 i_4+1} }{(1-x)^2} \Big) 
\nonumber
\\*
&= \frac{1}{(1-x)^2(1-x^2)(1-x^3)} \;.
\end{align}
One can check that any order of scales $i_j$ leads to a convergent
sum. 

On the other hand,
\begin{align}
\parbox{50mm}{\begin{picture}(40,32)
\put(0,0){\epsfig{file=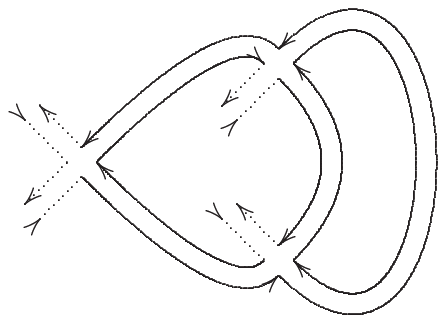,bb=71 579 192 669}}
\put(12,26){\mbox{\footnotesize$1$}}
\put(12,5){\mbox{\footnotesize$2$}}
\put(27,16){\mbox{\footnotesize$3$}}
\put(43,16){\mbox{\footnotesize$4$}}
\end{picture}}
\quad \mapsto \quad 
\parbox{40mm}{\begin{picture}(40,30)
\put(0,0){\epsfig{file=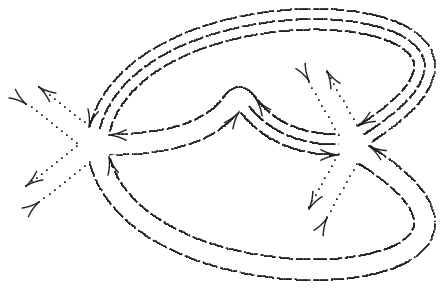,bb=71 591 191 672}}
\put(12,26){\mbox{\footnotesize$1$}}
\put(12,1){\mbox{\footnotesize$2$}}
\put(27,17){\mbox{\footnotesize$3$}}
\put(14,17){\mbox{\footnotesize$4$}}
\end{picture}}
\end{align}
The tree in the dual graph corresponds to $i_1\les i_2$ and $i_3\le
i_4$ and $i_1\les i_4$. For the particular choice $i_1\les i_2\les i_3\les
i_4$ of the scale attribution we have to compute according to
(\ref{AG-2}) the following sum, taking into account that the volume
factor from the summation over the main reference index is associated
to $i_3$:
\begin{align}
\sum_{i_4=0}^{\Lambda}\sum_{i_3=0}^{i_4}\sum_{i_2=0}^{i_3}
\sum_{i_1=0}^{i_2} x^{i_1+i_2-i_3+i_4}
&= \sum_{i_4=0}^{\infty}\sum_{i_3=0}^{i_4}\sum_{i_2=0}^{i_3} 
\frac{x^{i_2}-x^{2 i_2+1}}{1-x} x^{-i_3+i_4}
\nonumber
\\*
&= \sum_{i_4=0}^{\Lambda}\sum_{i_3=0}^{i_4}\Big( 
\frac{x^{-i_3}-x(1+x) + x^{3+i_3}}{(1-x)^2(1+x)} \Big) x^{i_4}
\nonumber
\\*
&= \sum_{i_4=0}^{\Lambda}\sum_{i_3=0}^{i_4}\Big( 
\frac{1-2x^{i_4+1} +2x^{i_4+3} -x^{2i_4+4}}{(1-x)^3(1+x)}
-\frac{i_4\,x^{i_4+1}}{(1-x)^2} \Big)
\nonumber
\\*
&= \frac{\Lambda}{(1-x)^2(1-x^2)} 
+ \frac{1-2 x +4x^2}{(1-x)^2(1-x^2)^2}
\nonumber
\\
&
+  \frac{\Lambda\,x^{\Lambda+2}}{(1-x)^3}
+ \frac{3 x^{\Lambda+2} + 4 x^{\Lambda+3}
- x^{\Lambda+4} - 2 x^{\Lambda+5}
+ x^{2\Lambda+6} }{(1-x)^2(1-x^2)^2}\;.
\end{align}
Thus, although the graph is \emph{superficially convergent}, there is
a \emph{subdivergence} given by the subgraph made of propagators 3 and
4, which leads to a divergent sum over the scale attribution.
Therefore, the subdivergence must be treated first by Taylor
subtraction at vanishing momenta in the same way as in
\cite{Grosse:2004yu}.

\end{appendix}


\begin{thebibliography}{99}

\bibitem{Schomerus:1999ug}
V.~Schomerus,
``D-branes and deformation quantization,''
JHEP {\bf 9906} (1999) 030
[arXiv:hep-th/9903205].

\bibitem{Seiberg:1999vs}
N.~Seiberg and E.~Witten,
``String theory and noncommutative geometry,''
JHEP {\bf 9909} (1999) 032
[arXiv:hep-th/9908142].

\bibitem{Minwalla:1999px}
S.~Minwalla, M.~Van Raamsdonk and N.~Seiberg,
``Noncommutative perturbative dynamics,''
JHEP {\bf 0002} (2000) 020
[arXiv:hep-th/9912072].

\bibitem{Chepelev:2000hm} I.~Chepelev and R.~Roiban, ``Convergence
  theorem for non-commutative Feynman graphs and renormalization,''
  JHEP {\bf 0103} (2001) 001 [arXiv:hep-th/0008090].

\bibitem{Grosse:2004yu} H.~Grosse and R.~Wulkenhaar, ``Renormalization
  of $\phi^4$-theory on noncommutative $\mathds{R}^4$ in the matrix
  base,'' arXiv:hep-th/0401128, to appear in Commun.\ Math.\ Phys.

\bibitem{Wilson:1973jj}
K.~G.~Wilson and J.~B.~Kogut,
``The renormalization group and the epsilon expansion,''
Phys.\ Rept.\  {\bf 12} (1974) 75.

\bibitem{Polchinski:1983gv}
J.~Polchinski,
``Renormalization and effective Lagrangians,''
Nucl.\ Phys.\ B {\bf 231} (1984) 269.

\bibitem{Grosse:2003aj}
H.~Grosse and R.~Wulkenhaar,
``Power-counting theorem for non-local matrix models and renormalization,''
arXiv:hep-th/0305066, to appear in Commun.\ Math.\ Phys.

\bibitem{Langmann:2002cc}
E.~Langmann and R.~J.~Szabo,
``Duality in scalar field theory on noncommutative phase spaces,''
Phys.\ Lett.\ B {\bf 533} (2002) 168
[arXiv:hep-th/0202039].

\bibitem{Rivasseau:1991ub}
V.~Rivasseau, ``From perturbative to constructive renormalization,''
Princeton Univ.\ Press, Princeton, 1991.

\bibitem{Grosse:2004by}
H.~Grosse and R.~Wulkenhaar,
``The $\beta$-function in duality-covariant noncommutative $\phi^4$-theory,''
Eur.\ Phys.\ J.\ C {\bf 35} (2004) 277 [arXiv:hep-th/0402093].

\bibitem{Langmann:2003if}
E.~Langmann, R.~J.~Szabo and K.~Zarembo,
``Exact solution of quantum field theory on noncommutative phase spaces,''
JHEP {\bf 0401} (2004) 017
[arXiv:hep-th/0308043].

\bibitem{Grosse:2004ik} H.~Grosse and R.~Wulkenhaar, ``Renormalization
  of $\phi^4$ theory on noncommutative $\mathbb{R}^4$ to all orders,''
  arXiv:hep-th/0403232, to appear in Lett.\ Math.\ Phys.


\bibitem{Rivasseau:2004az}
V.~Rivasseau and F.~Vignes-Tourneret,
``Non-commutative renormalization,''
arXiv:hep-th/0409312.


\end{thebibliography}
\end{document}